\documentclass[preprint, authoryear, 12pt]{elsarticle}
\usepackage{subfigure}
\usepackage{url}
\usepackage{graphicx}

\begin{document}

\title{
{Systematic variations in divergence angle}
}

\author{Takuya Okabe}
\address{
Faculty of Engineering, Shizuoka University, 3-5-1 Johoku, 
Hamamatsu 432-8561,Japan}


\begin{abstract}
Practical methods for quantitative analysis of  radial and angular
 coordinates of leafy organs of vascular plants  are presented and
 applied to published phyllotactic patterns of various real systems from
 young leaves on a shoot tip to florets on a flower head.  The constancy
 of divergence angle is borne out  with accuracy of less than a degree.
 It is shown that apparent fluctuations in divergence angle are in large
 part systematic variations caused by the invalid assumption of a fixed
 center and/or by secondary deformations,  while random fluctuations are
 of minor importance. 
\end{abstract}


\maketitle

\section{Introduction}
It has long since been recognized that 
divergence angles between successive 
leafy organs 
of vascular plants  
are accurately regulated 
at one of special constant values. 
Deviations from the the constant angle are normally so small that 
this botanical phenomenon, phyllotaxis,  
should rather be regarded 
as a genuine subject of exact science.
Besides the angular regularity, 
radial coordinates also exhibit mathematical regularity.

In a polar coordinate system, 
position of the $n$-th leaf is specified by 
the radial and angular coordinates $(r_n,\theta_n)$. 
The angular regularity is expressed by the equation 
\begin{equation}
 \theta_n = 
n d,  
\label{thetan=nd}
\end{equation}
where $d$ is a divergence angle. 
A spiral pattern is made when 
the radial component $r_n$ is a monotonic function of $n$, i.e., 
$r_n$ preserves the order of the leaf index $n$. 
Particularly important is 
a logarithmic spiral given by 
\begin{equation}
 r_n=a^n, 
\label{rn=an}
\end{equation}
where $a$ is a constant, called the plastochron ratio. 
Leaf primordia at a shoot tip appear to be arranged on a logarithmic spiral.
Eq.~(\ref{rn=an}) is a solution of 
the differential equation of exponential growth,  
\begin{equation}
\frac{{\rm d}}{{\rm d} n} \log  r_n = \log a. 
\label{ddnlogrn=loga} 
\end{equation}
Thus, 
the logarithmic spiral is characterized by the constant growth rate 
$\log a$. 
%
On the other side, a power-law spiral given by 
\begin{equation}
 r_n= \sqrt{n}
\label{rn=Asqrtn}
\end{equation}
has been often used as 
a mathematical model of packed seeds on a sunflower head
(\cite{vogel79,ridley82a,ropl84}). 
The equation (\ref{rn=Asqrtn}) is expected when all seeds
have equal areas, i.e., for the area per seed, 
\begin{equation}
\frac{{\rm d}}{{\rm d} n} (\pi r_n^2)={\rm const}. 
\label{ddnpirn2} 
\end{equation}

In mathematical studies, the relations 
(\ref{thetan=nd}), (\ref{rn=an}) and (\ref{rn=Asqrtn})
are often accepted 
without inquiring 
their empirical basis.  
%
From a physical standpoint, 
the validity of the mathematical relations has to be checked 
against experiments at all events. 
While embarking on quantitative assessment, however, 
we are confronted with a serious methodological problem. 
First of all, 
a center of the pattern, the origin of the polar coordinate system, 
must be located.  
Unfortunately,  it is not always possible to locate 
a fixed center  properly and objectively.
%
The uncertainty of the central position 
becomes a fundamental obstacle to assessing 
the empirical relations in a quantitative manner. 

Quantitative analysis of phyllotaxis has not been made until recently.    
%
%
\cite{me77} used a trigonometric method for evaluating divergence angle
from positions of three successive leaves.  
Their three-point method 
is based on the exponential growth (\ref{rn=an}), 
and it has been developed by \cite{meicenheimer86} and \cite{hotton03}. 
On the other side, \cite{rutishauser98} has estimated the plastochron ratio $a$ 
from the geometrical mean $({r_m}/{r_n})^{1/(m-n)}$ evaluated for selected leaves. 
The exponential growth is verified by the observation that the mean is nearly constant. 
%
\cite{rrb91} has assessed the validity of (\ref{rn=Asqrtn}) 
for real sunflowers by indirect means of evaluating 
the Euclidean distance between successive seeds,  
$ D_n^2=r_n^2+r_{n+1}^2-2r_nr_{n+1}\cos (\theta_n-\theta_{n+1})$. 
This method has a merit of being independent of the choice of the center
position. 
They have attempted parameter fittings with various functional forms of $r_n$. 
\cite{mkzm98} has investigated computational methods 
by regarding the center of gravity 
as the center of the pattern. 
\cite{hotton03} has proposed to use 
the minimal variation center that minimizes 
the standard deviation of divergence angles.


The main aim of this paper is to present practical methods 
for evaluating radial and angular coordinates of phyllotactic units 
without assuming any specific functional form of $r_n$ nor the existence
of a fixed center of the pattern.  
In Sec.~\ref{sec:floatingcenter}, a floating center of divergence 
is defined by positions of four consecutive leaves. 
In Sec.~\ref{sec:floatingcenterP}, 
a floating center of parastichy is defined by four nearby points
forming a unit cell of a parastichy lattice.  
%
For this purpose, 
a systematic index system for leaves of a multijugate pattern is
proposed in Sec.~\ref{sec:whorled}.     
In Sec.~\ref{sec:deformation}, effects of uniform deformation are considered. 
In Sec.~\ref{sec:results}, 
presented methods are 
applied to 
 real representative systems, which are arbitrarily chosen from previous works 
(\cite{me77, rrb91,rutishauser98,hjwzagd06}). 
In \ref{sec:logarithmicspiral}, mathematical characteristics of 
the logarithmic spiral pattern are examined carefully.  
To assist in interpreting the results in Sec.~\ref{sec:results}, 
general mathematical formulas for the plastochron ratio $a$ of systems with
orthogonal parastichies are derived.  
In \ref{how2number}, 
a practical method for numbering 
packed florets of a high phyllotaxis pattern is presented. 
The method is based on a number theoretic algorithm. 
To analyze phyllotaxis of Asteraceae,
\cite{hjwzagd06} have 
%
introduced a {\it primordia front}
 based on the assumption that the pattern has a definite center 
and the radius $r_n$ from the center is a monotonic function of $n$. 
%
%
%
In \ref{how2number}, 
a similar concept, a front circle on a parastichy lattice, 
is introduced without referring to the center.

\section{Floating center}

\subsection{Floating center of divergence}
\label{sec:floatingcenter}

For the purpose of this paper, it is suffice to investigate 
phyllotactic patterns projected on a plane. 
For three-dimensional analysis,
see \cite{hjwzagd06} and references therein. 
An effect due to a slight inclination of the normal axis is discussed in Sec.~\ref{sec:deformation}. 

\begin{figure}[t]
\begin{center}
\includegraphics[width=0.45\textwidth]{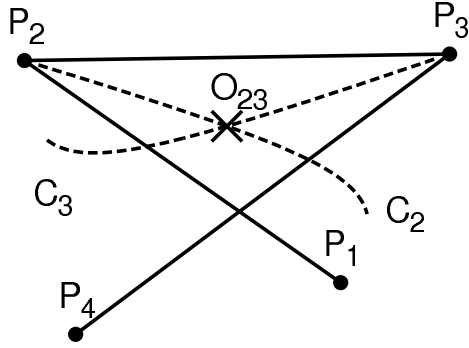}
\caption{
Given four consecutive points, $P_1$, $P_2$, $P_3$ and $P_4$, 
a center of divergence $O_{23}$ is defined such that $\protect\angle P_1 O_{23} P_2=
\protect\angle P_2 O_{23} P_3 =\protect\angle P_3 O_{23} P_4$. 
A dashed curve $C_2$ is a locus of the point $O_2$ satisfying 
$\protect\angle P_1 O_{2} P_2= \protect\angle P_2 O_{2} P_3$. 
Similarly, a dashed curve $C_3$ is drawn by 
the point $O_3$ satisfying $\protect\angle P_2 O_{3} P_3= \protect\angle P_3 O_{3} P_4$.  
The point where $C_2$ and $C_3$ intersect is $O_{23}$, 
the floating center of divergence.
\label{COD}
}
\end{center}
\end{figure}
Consider four consecutive points $P_1$, $P_2$, $P_3$ and $P_4$ 
 in the same geometric plane. 
Given the first three points $P_1$, $P_2$ and $P_3$, 
the point $O_2$ satisfying 
\begin{equation}
 \angle P_1O_2P_2= \angle P_2O_2P_3
\label{p1o2p2}
\end{equation}
defines a curve $C_2$ (Fig.~\ref{COD}). 
When $P_2P_1=P_2P_3$, or the triangle $P_1P_2P_3$ is isosceles,  
$C_2$ bisects the vertex angle $\angle P_1P_2P_3$. 
Similarly, for the three points $P_2$, $P_3$ and $P_4$, 
we obtain a curve $C_3$ traced by the point $O_3$ satisfying  
\begin{equation}
 \angle P_2O_3P_3= \angle P_3O_3P_4. 
\label{p2o2p3}
\end{equation}
The floating center of divergence $O_{23}$ is defined as
the crossing point of the two curves $C_2$ and $C_3$. 
By definition, 
\[
\angle P_1 O_{23} P_2=
\angle P_2 O_{23} P_3
=\angle P_3 O_{23} P_4. 
\]
Thus,  
the three angles spanned by the four consecutive points
define a divergence angle,  
\begin{equation}
 d_2\equiv \angle P_2 O_{23} P_3. 
\end{equation}
Distance from $O_{23}$ is denoted as
\[
 r_2\equiv |\overrightarrow{O_{23}P_2}|. 
\]
In general, 
the divergence angle $d_n$ and 
the radius $r_n$ are evaluated 
from four consecutive points $P_{n-1}$, $P_{n}$, $P_{n+1}$ and $P_{n+2}$; 
\begin{equation}
 d_n\equiv \angle P_n O_{nn+1} P_{n+1},  
\label{dn}
\end{equation}
\begin{equation}
 r_n\equiv |\overrightarrow{O_{nn+1}P_n}|. 
\end{equation}
These quantities are compared with 
\begin{eqnarray}
d_n^{(0)} &=&\angle P_n O_{0} P_{n+1}, 
\label{dn(0)}
\end{eqnarray}
and 
\begin{equation}
 r_n^{(0)}=|\overrightarrow{O_{0}P_n}|,  
\label{rn(0)}
\end{equation}
which are defined in terms of an arbitrarily chosen center $O_{0}$.

Given coordinates of the four points $P_n$ $(n=1,2,3,4)$,  
the position of the floating center $O_{23}$ is 
parametrically represented by means of two parameters $X$ and $Y$; 
\begin{equation}
 \overrightarrow{P_2O_{23}}=X\overrightarrow{P_2P_3}+Y\overrightarrow{P_2P_1}. 
\label{P2O23=X}
\end{equation}
The two parameters are determined by the two equations
(\ref{p1o2p2}) and (\ref{p2o2p3}), namely  
\begin{equation}
\frac{
|\overrightarrow{O_{23}P_1}|^2+|\overrightarrow{O_{23}P_2}|^2-
|\overrightarrow{P_1P_2}|
}{ |\overrightarrow{O_{23}P_1}|
|\overrightarrow{O_{23}P_2}|
}
=
\frac{
|\overrightarrow{O_{23}P_2}|^2+|\overrightarrow{O_{23}P_3}|^2-
|\overrightarrow{P_2P_3}|
}{ |\overrightarrow{O_{23}P_2}|
|\overrightarrow{O_{23}P_3}|
}, 
\label{O23P1}
\end{equation}
and 
\begin{equation}
\frac{
|\overrightarrow{O_{23}P_2}|^2+|\overrightarrow{O_{23}P_3}|^2-
|\overrightarrow{P_2P_3}|
}{ |\overrightarrow{O_{23}P_2}|
|\overrightarrow{O_{23}P_3}|
}
=
\frac{
|\overrightarrow{O_{23}P_3}|^2+|\overrightarrow{O_{23}P_4}|^2-
|\overrightarrow{P_3P_4}|
}{ |\overrightarrow{O_{23}P_3}|
|\overrightarrow{O_{23}P_4}|
} 
\label{O23P2}
\end{equation}
by the cosine formula in trigonometry. 

For example, consider vertex points $P_n$ ($n=1,2,3,4,5$) of a regular pentacle, 
whose Cartesian coordinates are given by 
\(
(x_n, y_n)=(\cos nd, \sin nd ), 
\)
where $d=2\pi \alpha$ in radians and $\alpha=\frac{2}{5}$. 
Let the center $O_{23}$ of the four points $P_n$ ($n=1,2,3,4$) be
represented as (\ref{P2O23=X}). 
For this special case, 
it is not difficult to find the solution $X=Y=\frac{1}{2+\tau}\simeq
0.2764$ by geometrical considerations.
The golden ratio
\begin{equation}
\tau=\frac{\sqrt{5}+1}{2}
\label{tau}
\end{equation}
is 
the irrational number quintessential to the phenomenon of  phyllotaxis.  
The estimate of  $X\simeq Y\sim 0.3$ can be used as an initial guess
for the numerical search of solutions in general cases. 

The manner in which 
the floating center floats around 
may be illustrated by means of 
a line segment connecting the floating center and 
the middle point of the middle two leaves defining the center, 
that is, the line segment $O_{23}M_{23}$,  
where $M_{23}$ is the middle point of $P_{2}$ and $P_{3}$.
Let us call 
a graph of the line segments  {\it a divergence diagram}. 
If the center is fixed in space, 
all the line segments radiate from the fixed center.
If it is not, the lines cross with each other. 
See Fig.~\ref{spiral}, for instance.

%

\begin{figure}[t]
\begin{center}
\includegraphics[width=0.4\textwidth]{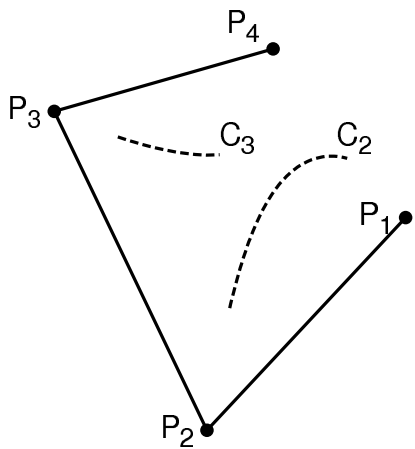}
\caption{
Four points $P_i$ ($i=1,2,3,4$) from a deformed Lucas pattern 
for which two curves $C_2$ and $C_3$ 
as defined in Fig.~\ref{COD} fail to cross. 
Thus, it may happen that the center of divergence is not defined 
when divergence angle is small  and variable. 
}
\label{lucas}
\end{center}
\end{figure}

When a pattern with small divergence angles is deformed significantly,
it may happen that 
no center is defined because the two curves $C_2$ and $C_3$ do not cross
(Fig.~\ref{lucas}). 
Even in such a case, a center can be defined formally by selecting four
points properly.  
%
Here 
it is remarked only that 
{a pattern consisting of more than three points may not
have a definite center.}
In other words, it is very special for 
numerous leaves comprising a phyllotactic pattern to have a unique, fixed center. 
Assuming the fixed center is not at all a trivial matter. 

%
%
%

%


\subsection{Floating center of parastichy}
\label{sec:floatingcenterP}

\subsubsection{Spiral phyllotaxis}

\begin{figure}[t]
\begin{center}
\includegraphics[width=0.45\textwidth]{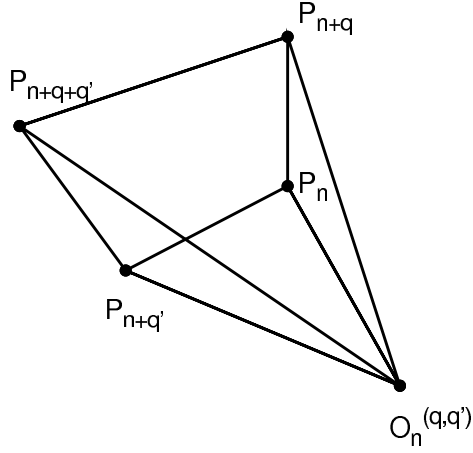}
\caption{
From four nearby points $P_n$, $P_{n+q}$, $P_{n+q'}$ and $P_{n+q+q'}$
making a tetragonal cell of a lattice of $q$ and $q'$ parastichies, 
a center of parastichy $O_n^{(q,q')}$ is determined by 
(\ref{q'angle P_nO_n}) and (\ref{anglePnOnqq'}).  
By definition, four angles subtended by four sides of the tetragon 
give a unique divergence angle against $O_n^{(q,q')}$, 
which is called the floating center of parastichy.  
\label{fig:cop}
}
\end{center}
\end{figure}
For a spiral pattern with one leave per each node, 
leaves are orderly indexed by an integer $n$, called the plastochron. 
Four points for a center need not be consecutive in plastochron. 
A divergence angle is evaluated from 
four vertex points $P_n$, $P_{n+q}$, $P_{n+q'}$ and $P_{n+q+q'}$ 
of a unit cell of a parastichy lattice, 
where $q$ and $q'$ are opposite parastichy numbers. 
Parastichies of logarithmic spirals are discussed 
in \ref{sec:logarithmicspiral}, 
where it is shown that divergence angle $d=2\pi \alpha_0$ of an ideal pattern 
satisfies the inequalities (\ref{p'q'alpha0pq}) 
in terms of two auxiliary integers $p$ and $p'$ satisfying $pq'-p'q=1$, (\ref{pq'-p'q=1}). 
With this in mind, 
a {\it center of parastichy} $O_n^{(q,q')}$ is defined 
such that 
\begin{equation}
q' \angle P_{n}O_n^{(q,q')} P_{n+q}
+
q
\angle P_{n}
O_n^{(q,q')}
 P_{n+q'}=2\pi, 
\label{q'angle P_nO_n}
\end{equation}
and 
\begin{equation}
\angle P_{n}
O_n^{(q,q')}
 P_{n+q}
=
\angle P_{n+q'}
O_n^{(q,q')}
 P_{n+q+q'}. 
\label{anglePnOnqq'}
\end{equation}
where $\angle P_{n}O_n^{(q,q')} P_{n+q}$ signifies
the angle subtended by line segments $O_n^{(q,q')} P_{n}$ and $O_n^{(q,q')} P_{n+q}$. 
See Fig.~\ref{fig:cop}. 
Then, four angles made by the four points 
give a divergence angle; 
\begin{eqnarray}
 d_n^{(q,q')}&= &
\left(2\pi p- 
\angle P_{n}
O_n^{(q,q')}
 P_{n+q}
\right)/q
\nonumber\\&=&
\left(2\pi p- 
\angle P_{n+q'}
O_n^{(q,q')}
 P_{n+q+q'}
\right)/q
\nonumber\\&=&
\left(
\angle P_{n}
O_n^{(q,q')}
 P_{n+q'}
+2\pi p'
\right)/q'
\nonumber\\&=&
\left(
\angle P_{n+q}
O_n^{(q,q')}
 P_{n+q+q'}
+2\pi p'
\right)/q'. 
\label{d_n^{(q,q')}}
\end{eqnarray}
The rationale behind this definition may be understood by expressing the first equation as 
$\angle P_{n}O_n^{(q,q')} P_{n+q}=2\pi p-  d_n^{(q,q')} q $, which 
 is to be compared with the denominator of (\ref{b}). 
This is the net angle between the two successive points 
along a $q$ parastichy $P_{n}$ and $P_{n+q}$ 
when divergence angles between the two points are equal to $d_n^{(q,q')}$. 
The difference in the signs 
in front of geometrical angles like $\angle P_{n}O_n^{(q,q')} P_{n+q}$
in (\ref{d_n^{(q,q')}})
is due to 
the convention that 
the angles including $d_n^{(q,q')}$ are regarded as positive quantities.

Coordinates of the center $O_n^{(q,q')}$ are 
represented in terms of 
two parameters $r$ and $\theta$ as  
\begin{equation}
 \overrightarrow{
P_{n}
O_n^{(q,q')}
}
=
r\cos\theta 
 \overrightarrow{
P_{n+q}P_{n}
}
+
r\sin\theta 
 \overrightarrow{
P_{n+q'}P_{n}
}. 
\label{PnOnqq'=}
\end{equation}
Accordingly, (\ref{q'angle P_nO_n}) and  (\ref{anglePnOnqq'}) are
regarded as the defining equations for $r$ and $\theta$.  
As is clear from Fig.~\ref{fig:cop}, 
it is generally assumed that $r>0$ and $0< \theta < \frac{\pi}{2}$. 
By regarding $\theta$ as a given constant, 
the first equation (\ref{q'angle P_nO_n}) determines $r$ for the
given value of $\theta$, or $r$ is determined as a function of $\theta$. 
Then, the parameter $\theta$  is fixed by the second equation (\ref{anglePnOnqq'}). 
Thus, the center $O_n^{(q,q')}$ 
for given $P_n$, $P_{n+q}$, $P_{n+q'}$ and $P_{n+q+q'}$ is fixed numerically.  
The equations depend not only on the coordinates of the four points, 
but also on the parastichy numbers $q$ and $q'$, 
while either $p$ or $p'$ is needed 
to evaluate the divergence angle in (\ref{d_n^{(q,q')}}). 

A caveat: 
In the above, it is assumed that $P_{n}$ with the smallest index $n$
is positioned nearest to the center (Fig.~\ref{fig:cop}). 
In the opposite convention,
$P_{n}$ is farther from the center than $P_{n+q+q'}$. 
To adapt the above results to this case, 
the four points $P_n$, $P_{n+q}$, $P_{n+q'}$ and $P_{n+q+q'}$
should be replaced by 
$P_{n+q+q'}$, $P_{n+q'}$, $P_{n+q}$ and $P_n$, respectively. 
%
%
%
%
A set of equations (\ref{q'angle P_nO_n}), (\ref{anglePnOnqq'}), 
and (\ref{d_n^{(q,q')}}) 
is invariant by this replacement, whereas (\ref{PnOnqq'=}) should be read as
\begin{equation}
 \overrightarrow{P_{n+q+q'}O_n^{(q,q')}}
=
r\sin\theta 
 \overrightarrow{
P_{n+q}P_{n+q+q'}
}
+
r\cos\theta 
 \overrightarrow{
P_{n+q'}P_{n+q+q'}
}. 
\end{equation}

In the biological literature, leaves on a stem are numbered in the order
of appearance, while primordia on an apex are counted in the
opposite order.

While $d_n$ in (\ref{dn}) based on four distant points 
is insensitive to individual displacements of the points, 
$d_n^{(q,q')}$ in (\ref{d_n^{(q,q')}}) based on nearby points 
is insensitive to a collective displacement of the points. 
The latter has a merit of wider applicability in practice.  
It is stressed again that 
parastichies generally do not have a definite center; 
the center $O_n^{(q,q')}$ thus determined 
depends on the plastochron $n$ and the parastichy
pair $(q,q')$. 

\subsubsection{Multijugate phyllotaxis}
\label{sec:whorled}

A multijugate pattern 
bears more than one leaves at each node. 
There has been no established way of systematize multijugate leaves, 
particularly because a special index to distinguish leaves at a node is lacking.  
Therefore, in the first place, 
 a systematic index system  for a multijugate pattern is proposed below. 
Then, the center of parastichy is defined similarly as 
in the last subsection.  

A $J$-jugate pattern has $J$ leaves at a node. 
There are $J$ fundamental spirals correspondingly. 
Each of $J$ leaves at the $n$-th node is specified 
by polar coordinates $(r_{n,j},\theta_{n,j})$, 
where the jugacy index $j=0,1,2,\cdots, J-1$ is introduced.  
Any leaf may be chosen as 
the origin $(n,j)=(0,0)$ of the index system. 
Along with the plastochron $n$, the jugacy index $j$ is set in order 
in the direction of the fundamental spirals. 
The leaf with the index $(n,j)$ is referred to by the symbol $P_n^j$, or simply denoted as $n^j$.
For instance, see Fig.~\ref{spiralrrb3} for a trijugate system with $J=3$. 
For convenience' sake, 
the value range of the jugacy index $j$ is extended to 
all integers; 
the coordinates $(r_{n,j},\theta_{n,j})$ 
are regarded as periodic in $j$ with period of $J$, 
namely $(r_{nj+J},\theta_{nj+J})=(r_{n,j},\theta_{n,j})$ 
for all $n$ and $j$. 
Accordingly, divergence angle
\begin{equation}
d_n^{j}=\theta_{n+1,j}- \theta_{n,j}
\label{dnj=|}
\end{equation}
is periodic in $j$. 
As implied by the notation, $d_n^{j}$ depends on the leaf index $(n,j)$, 
whereas it becomes constant, $d_n^{j}=d$, for an ideal pattern. 

To put it concretely,  let us consider an ideal 
multijugate system spiraling in the positive direction, i.e., $ d>0$. 
A pattern with a negative angle $d <0$ is the mirror image of the positive counterpart. 
In an ideal $J$-jugate pattern, 
the radial coordinate $r_n^j$ is independent of the jugacy index $j$,
\[
r_{n,j}=r_n,  
\]
whereas the angular coordinate is given by 
\begin{equation}
 \theta_{n,j} = 
2\pi 
\left(
n \alpha_0 + 
\frac{j}{J} 
\right),  
\label{thetanj=}
\end{equation}
where the divergence angle $d=2\pi\alpha_0$ is represented with 
a dimensionless parameter $\alpha_0$. 
Without loss of generality, $0<J\alpha_0< \frac{1}{2}$. 
%
For multijugate patterns $J>1$, 
the other quantity of interest is displacement angle between neighboring leaves at a node, 
\begin{equation}
\Delta_n^{j}=\theta_{n, j+1}- \theta_{n, j} - \frac{2\pi}{J}. 
\label{Deltanj=thetanj+1}
\end{equation}
For the ideal pattern with (\ref{thetanj=}),  $\Delta_n^{j}=0$.  
In general, $\Delta_n^{j}$
fluctuates evenly about zero. 

In this index system, 
a parastichy is specified by a set of numbers $(q,j)$, 
increments of $(n,j)$. 
Conspicuous parastichies following nearby leaves are shown to  have $j=-p$, 
where $p$ is an integer near $J\alpha_0 q$ (cf. below (\ref{b})). 
Thus, the $q$ parastichy running through a leaf $P_n^i$ connects the points 
$P_n^i$, $P_{n+q}^{i-p}$, 
$P_{n+2q}^{i-2p}$, 
$P_{n+3q}^{i-3p}, \cdots$. 
For instance, the trijugate pattern in Fig.~\ref{spiralrrb3}
has $J=3$ and $(q,p)=(3,1)$, $(5,2)$, etc. 
Note that a 3-parastichy $1^0-4^{-1}-7^{-2}-10^{-3}-\cdots$ 
is equivalently represented as $1^0-4^2-7^1-10^0-\cdots$, 
as the jugacy superscript $j$ is understood modulo $J=3$
by the periodic extension prescribed above.

Having thus prepared, 
the method of the parastichy center in the last subsection is applied. 
The divergence angle $d_n^{j (q,q')}$ of a multijugate system is defined
by the four points $P_n^j$, 
$P_{n+q}^{j-p}$, 
$P_{n+q'}^{j-p'}$, $P_{n+q+q'}^{j-p-p'}$; 
%
(\ref{q'angle P_nO_n}),  (\ref{anglePnOnqq'}) 
and (\ref{d_n^{(q,q')}}) should read
\begin{equation}
q' \angle P_{n}^jO_n^{(q,q')} P_{n+q}^{j-p}
+
q
\angle P_{n}^j
O_n^{(q,q')}
 P_{n+q'}^{j-p'}=2\pi/J, 
\end{equation}
\begin{equation}
\angle P_{n}^j
O_n^{(q,q')}
 P_{n+q}^{j-p}
=
\angle P_{n+q'}^{j-p'}
O_n^{(q,q')}
 P_{n+q+q'}, 
\end{equation}
and
\begin{eqnarray}
 d_n^{j (q,q')}&= &
\left(
\frac{2\pi p}{J}- 
\angle P_{n}^j
O_n^{(q,q')}
 P_{n+q}^{j-p}
\right)/q
\nonumber\\&=&
\left(
\frac{2\pi p}{J}- 
\angle P_{n+q'}^{j-p'}
O_n^{(q,q')}
 P_{n+q+q'}^{j-p-p'}
\right)/q
\nonumber\\&=&
\left(
\angle P_{n}^j
O_n^{(q,q')}
 P_{n+q'}^{j-p'}
+\frac{2\pi p'}{J}
\right)/q'
\nonumber\\&=&
\left(
\angle P_{n+q}^{j-p}
O_n^{(q,q')}
 P_{n+q+q'}^{j-p-p'}
+\frac{2\pi p'}{J}
\right)/q'.
\end{eqnarray}

\section{Effect of uniform deformation}
\label{sec:deformation}

Consider a regular spiral pattern in which 
position of the $n$-th leaf is given by Cartesian coordinates $(x_n, y_n)$. 
The center $O_0$ of the polar coordinate system $(r_n^{(0)}, \theta_n^{(0)})$
is set at the origin. 
The azimuthal angle $\theta_n^{(0)}$ is measured from the $x$-axis.
As typical cases for a phyllotactic pattern to lose track of the center, 
two types of global deformation 
are considered. 
The deformation is 
characterized by a single parameter representing strain $C$.

Uniform deformation of the first order transforms $(x_n, y_n)$ as follows; 
\begin{eqnarray}
 x_n &\longrightarrow &
\left\{
\begin{array}{ll}
(1+C_1) x_n. &x_n\ge 0 \\
(1-C_1) x_n. &x_n<0  \\
\end{array}
\right. 
\nonumber\\
y_n &\longrightarrow&  y_n.  
\label{xnynC1}
\end{eqnarray}
Uniform deformation of the second order makes 
\begin{eqnarray}
 x_n &\longrightarrow  & (1-C_2) x_n, 
\nonumber\\ 
y_n &\longrightarrow  & y_n.
\label{xnynC2}
\end{eqnarray}
The magnitude of deformation is represented by 
constants $C_1$ and $C_2$, respectively, 
which are assumed significantly smaller than 1. 
In both cases,  the radial component 
$r_n^{(0)}\equiv \sqrt{x_n^2+y_n^2}$ and 
divergence angle $\theta_n^{(0)}-\theta_{n-1}^{(0)}$
suffer modulation periodic in the azimuthal angle $\theta_n^{(0)}$. 
The periods of modulation for the first and the second order deformation are 
360 and 180 degrees,  respectively.

The amplitude of modulation in the radius 
$r_n^{(0)} \pm \Delta r_n^{(0)}$ is given by 
\begin{equation}
\frac{\Delta r_n^{(0)}}{{r_n^{(0)}}} =C_1
\end{equation}
for the first order deformation, 
and 
\begin{equation}
\frac{\Delta r_n^{(0)}}{{r_n^{(0)}}} =\frac{C_2}{2}
\end{equation}
for the second order deformation. 

Amplitude of modulation in divergence angle, ${\Delta d}^{(0)}$,  
may be estimated as follows. 
Consider an isosceles triangle 
with the vertex angle $d^{(0)}$, the base of length $2L$, and the height $H$. 
Then,  
\[
 \tan \frac{d^{(0)}}{2} =\frac{L}{H}. 
\]
By the first order deformation, 
$d^{(0)}$ becomes  $d^{(0)} \pm \Delta d^{(0)}$ 
when the height $H$ parallel to the $x$ axis is modified to $(1 \mp C_1)H$. 
By the second order deformation, 
$d$ is modified  to $d^{(0)} \pm \Delta d^{(0)}$ 
as $(H,L)$ become $((1-C_2)H, L)$ and $(H, (1-C_2)L)$.  
In either case, 
\begin{equation}
  \tan \frac{d^{(0)} \pm\Delta d^{(0)}}{2} \simeq \frac{L}{H} (1\mp C), 
\end{equation}
where $C$ is $C_1$ or $C_2$. 
Accordingly, 
\begin{equation}
{\Delta d^{(0)}} \simeq  C {\sin d^{(0)} }.   
\end{equation}
In particular, 
\begin{equation}
{\Delta d^{(0)}} \simeq  38.7 C\ {\rm degrees}
\label{Deltadsimeq38.7}
\end{equation}
for $d^{(0)}=137.5$ degrees, 
and 
\begin{equation}
{\Delta d^{(0)}} \simeq  56.5 C\ {\rm degrees}
\label{Deltadsimeq565}
\end{equation}
for $d^{(0)}=99.5$ degrees. 

The first order deformation may be caused 
by a tendency toward or against the direction of the sun (\cite{kumazawa71}). 
The second order deformation may be caused by various reasons,  real or apparent. 
The most possible would be due to uniform compression. 
It applies also to the case where 
a pattern is observed from a direction oblique to the perpendicular direction. 
If the angle of inclination is denoted as $\varphi$ in radian measure, 
then $1-C_2=\cos \varphi$, or $C_2\simeq {\varphi^2}/{2}$.

%

\section{Results}
\label{sec:results}

\subsection{Example}

\begin{figure}
  \begin{center}
  \subfigure[]{
\includegraphics[width= .38\textwidth]{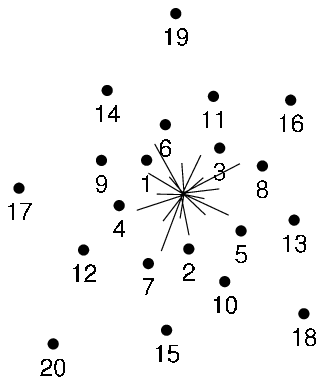}
    \label{spiralC1005}
 }
  \hfill
  \subfigure[]{
\includegraphics[width= .38\textwidth]{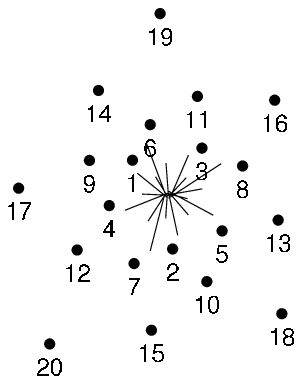}
    \label{spiralC005}
  }
  \end{center}
  \vspace{-0.5cm}
  \caption{ 
An ideal pattern of a logarithmic spiral with the plastochron ratio
 $a=1.073$ is compressed 
(a) by (\ref{xnynC1}) with $C_1=0.05$, and 
(b) by (\ref{xnynC2}) with $C_2=0.05$. 
Leaves are represented with points with integer index $n$. 
Radiating lines
represent $O_{nn+1}M_{nn+1}$, 
where $O_{nn+1}$ is the floating center of divergence 
for the four consecutive leaves from the $(n-1)$-th to $(n+2)$-th leaves and  
$M_{nn+1}$ is the middle point  of the $n$-th and $(n+1)$-th leaves. 
It is called a divergence diagram in Sec.~\ref{sec:floatingcenter}. 
}
  \label{spiral}
\end{figure}

\begin{figure}
\centering
  \subfigure[$d^{(0)}$ and $d$ against $n$]{
\includegraphics[width= .40\textwidth]{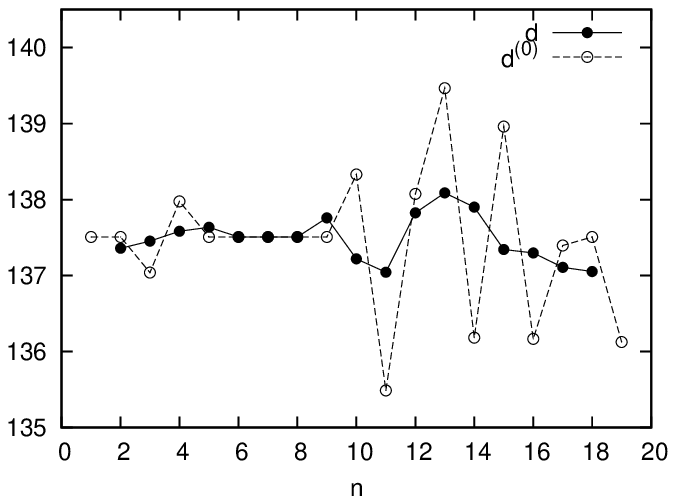}
    \label{sample1div}
 }
  \subfigure[$r^{(0)}$ and $r$ against $n$]{
\includegraphics[width= .40\textwidth]{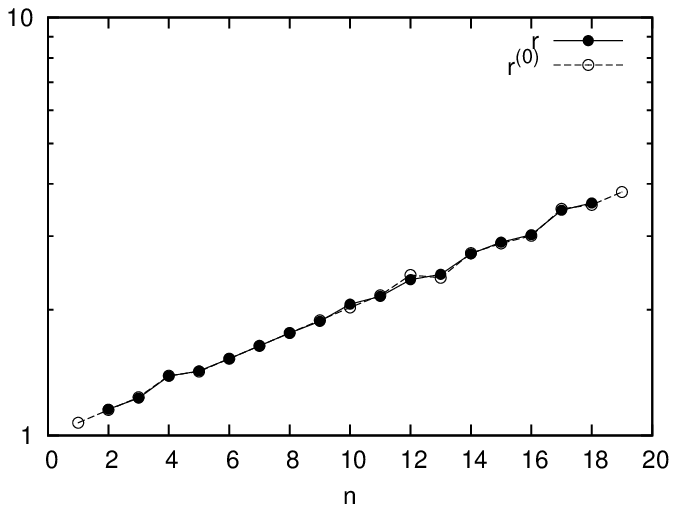}
  }
  \subfigure[$d^{(0)}$ and $d$ against $\theta$]{
\includegraphics[width=0.40 \textwidth]{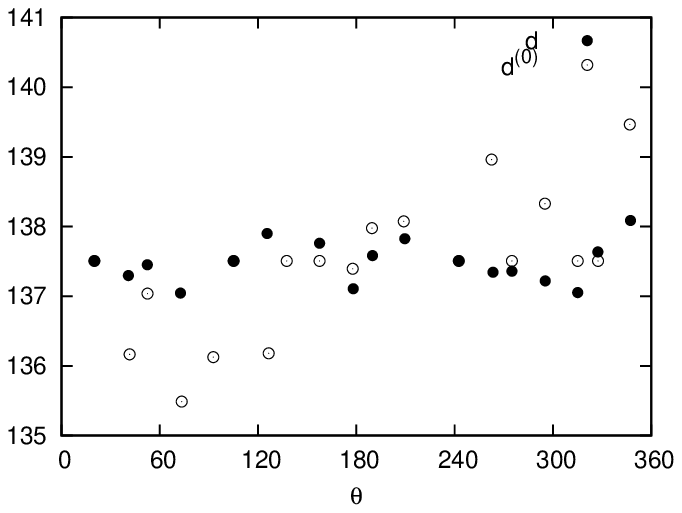}
    \label{sample1azim}
}
  \subfigure[$\Delta r^{(0)}/r^{(0)}$ and $\Delta r/r$ against $\theta$]{
\includegraphics[width=0.40 \textwidth]{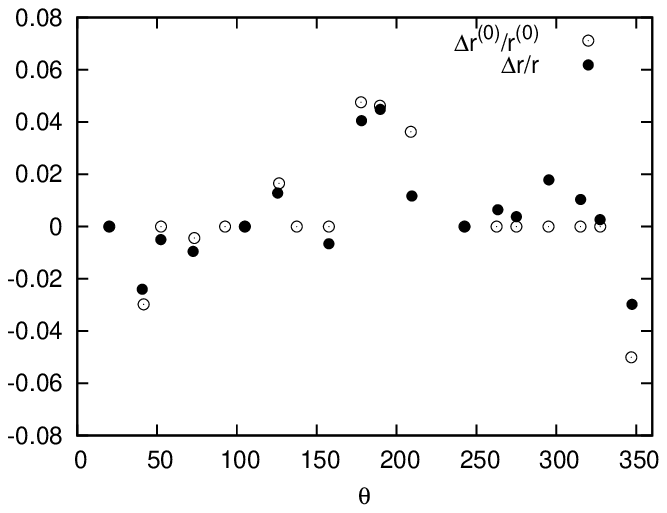}
    \label{sample1azimdelr}
}
  \caption{ 
Results for the deformed pattern in Fig.~\ref{spiralC1005}. 
$d^{(0)}$ and $r^{(0)}$ are divergence angle and radial coordinate 
with respect to the center of the original (undeformed) spiral pattern, 
while  $d$ and $r$ are divergence angle and radial coordinate 
with respect to the floating center of divergence. 
For the horizontal axis, $n$ is the leaf index, and 
$\theta$ is the polar angle, or azimuth.
$\Delta r^{(0)}$ and $\Delta r$ are 
changes in $r^{(0)}$ and $r$ by the uniform deformation for $C_1=0.05$. 
}
  \label{sample1divrad}
\end{figure}
\begin{figure}
\centering
  \subfigure[$d^{(0)}$ and $d$ against $n$]{
\includegraphics[width= .45\textwidth]{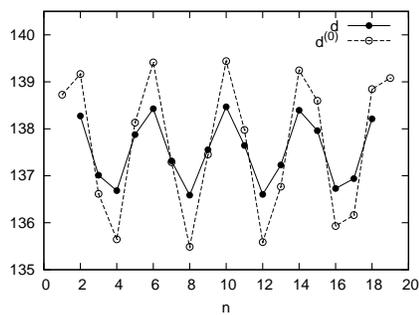}
 }
  \subfigure[$r^{(0)}$ and $r$ against $n$]{
\includegraphics[width= .45\textwidth]{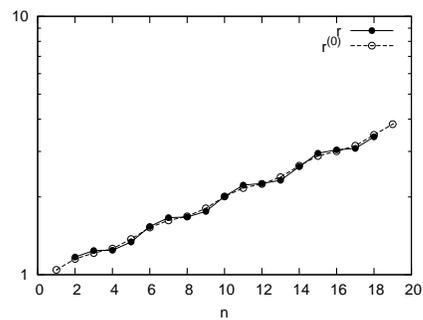}
  }
  \subfigure[$d^{(0)}$ and $d$ against $\theta$]{
\includegraphics[width=0.45 \textwidth]{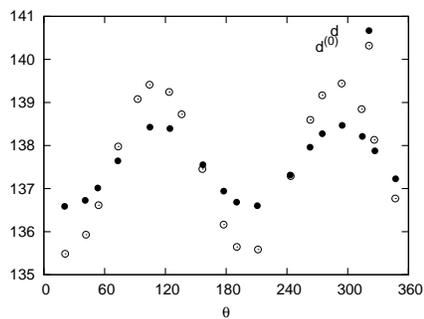}
    \label{sampleazim}
}
  \subfigure[$\Delta r^{(0)}/r^{(0)}$ and $\Delta r/r$ against $\theta$]{
\includegraphics[width=0.45 \textwidth]{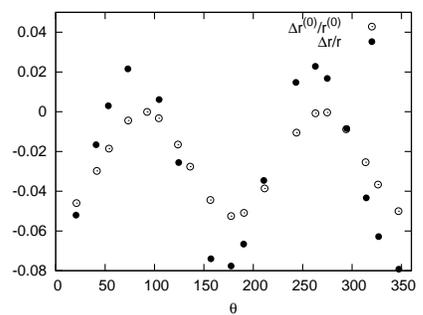}
    \label{sampleazimdelr}
}
  \caption{ 
Results for the deformed pattern in Fig.~\ref{spiralC005} ($C_2=0.05$); cf. Fig.~\ref{sample1divrad}. 
}
  \label{sampledivrad}
\end{figure}
\begin{figure}
  \begin{center}
  \subfigure[]{
\includegraphics[width= .45\textwidth]{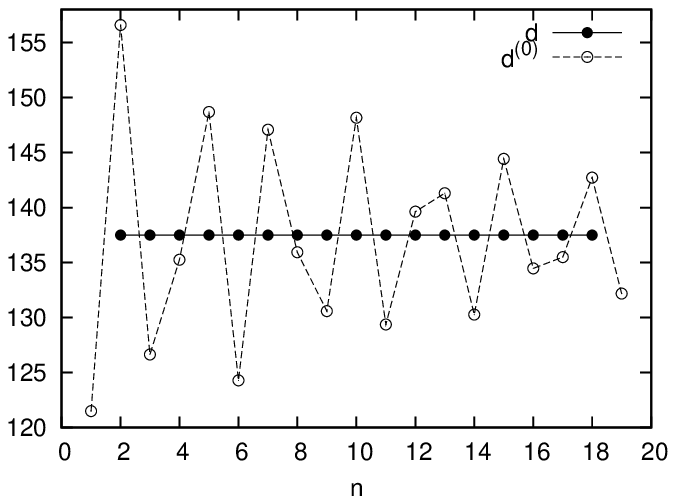}
 }
  \hfill
  \subfigure[]{
\includegraphics[width= .45\textwidth]{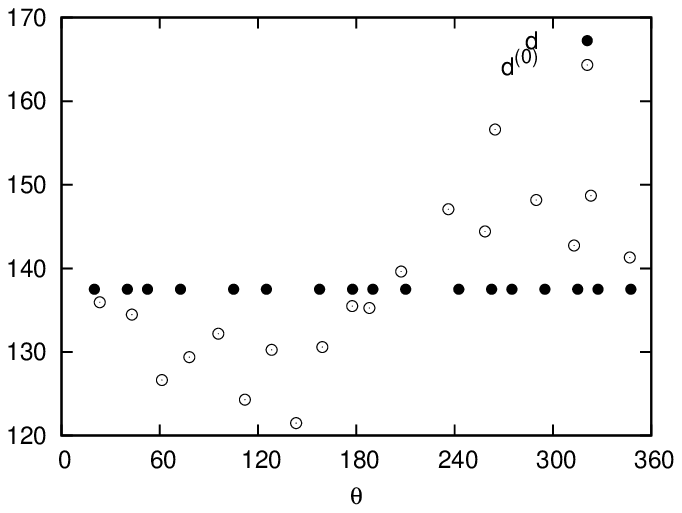}
    \label{azimsample2}
  }
  \end{center}
  \vspace{-0.5cm}
  \caption{ 
Divergence angles of
the original (not deformed) pattern of Fig.~\ref{spiral} are 
measured against a misplaced center $O_0$
set at the middle point of leaves 1 and leaf 5. 
The misplacement brings about apparent wild variations in $d_n^{(0)}$
 (open circles, left), 
which are correlated with azimuth $\theta_n$  (right). 
Divergence angle $d_n$ evaluated with 
the floating center of divergence (filled circles)
does not depend on the wrong center $O_0$; 
the original divergence angle is correctly retrieved without knowing the true center. 
}
  \label{azim_sample2}
\end{figure}
Consider 
a regular pattern with $d=2\pi/\tau^2$ (137.5 degrees) and $a=1.073$ for 
(\ref{thetan=nd}) and (\ref{rn=an}).
This is an orthogonal $(5,8)$ system corresponding to $i=5$ in (\ref{loga2pisqrt5tau2n+1}).  
By deforming the pattern according to 
(\ref{xnynC1}) with $C_1=0.05$ and (\ref{xnynC2}) with $C_2=0.05$, 
patterns shown in Figs.~\ref{spiralC1005} and \ref{spiralC005} are obtained, respectively. 
In Fig.~\ref{spiral}, 
line segments radiating from the center represent 
the divergence diagram explained at the end of Sec.~\ref{sec:floatingcenter}.
The fixed center $O_0$ 
for divergence angle $d^{(0)}$ and radius $r^{(0)}$ 
is set at the center of the original pattern. 
Divergence angles $d^{(0)}$, $d$  and radii $r^{(0)}$, $r$ 
for the pattern of Figs.~\ref{spiralC1005} 
and \ref{spiralC005} 
are shown in
Figs.~\ref{sample1divrad} 
and \ref{sampledivrad}, respectively. 
All quantities show characteristic variations. 
A type of variations as exhibited by $d^{(0)}$ in Fig.~\ref{sample1div}
is frequently met in real systems. 
Fig.~\ref{sample1azim} shows 
that the variations are correlated with azimuth $\theta_n$.
In general, 
deformation 
has a larger effect on the divergence angles than on the radii. 
Variations in the divergence angle $d$ by means of the floating center in Sec.~\ref{sec:floatingcenter} 
are considerably suppressed compared with those of $d^{(0)}$. 
A period of 180 degrees in Fig.~\ref{sampleazim} 
is the characteristic of 
uniform compression in one direction, namely
deformation of the second order.
Divergence angles facing the compressed direction increase apparently. 
For $C_2=0.05$,  (\ref{Deltadsimeq38.7}) gives $\Delta d\simeq 1.9$ degrees, 
which corresponds to the amplitudes of $d^{(0)}$ in Fig.~\ref{sampleazim} 
and $\Delta r^{(0)}/r^{(0)}$ in Fig.~\ref{sampleazimdelr}.
The amplitudes of $d$ and $\Delta r/r$  are suppressed and enhanced 
compared with $d^{(0)}$ and $\Delta r^{(0)}/r^{(0)}$, respectively.
The factors of suppression and enhancement are about the same value $\simeq 1.8$. 

Even if a pattern has a definite center, 
misidentification of the center causes apparently systematic variations
in $d^{(0)}$. 
Fig.~\ref{azim_sample2} is a result for 
the undistorted pattern $C_1=C_2=0$, 
whereas the middle point of 1 and 5 is chosen as the nominal center $O_0$. 
The divergence $d$ based on the floating center 
does not depend on the choice of $O_0$, whereas
the variations in $d^{(0)}$ exhibit a period of 360 degrees when plotted against the azimuth $\theta$. 
Thus, the misplacement of the center can be misinterpreted as the first order deformation. 
%

These results indicate that 
systematic features of variations may not be revealed unless they are plotted against the azimuth. 

\subsection{{Suaeda vera}}
\label{sec:suaedavera}
\begin{figure}
\centering
\subfigure[Position of leaves $(x_n, y_n)$]{
\includegraphics[height= .42\textwidth]{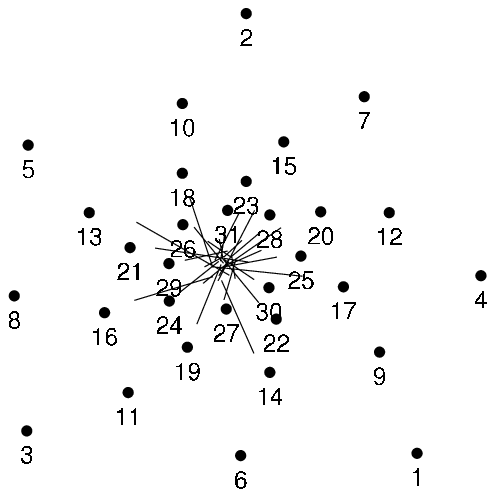}
\label{rutishauser:subfig1}
}
\subfigure[Divergence angles $d_n$ and $d^{(0)}_n$ against the leaf index $n$]{
\includegraphics[width= .45\textwidth]{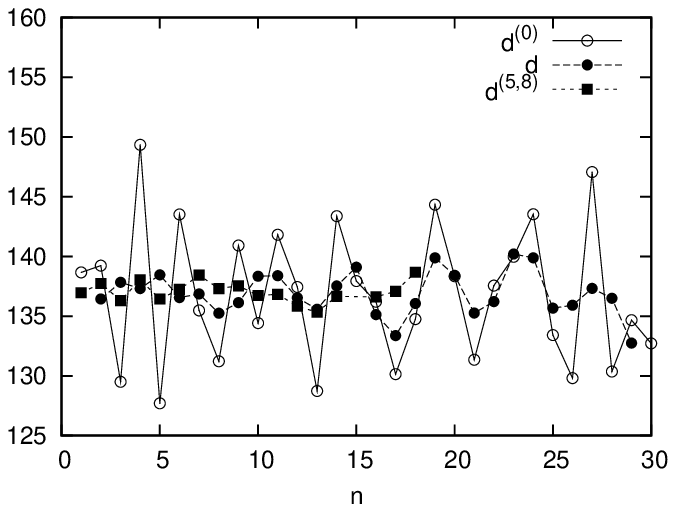}
\label{rutishauser:subfig2}
}
\subfigure[$d_n$ and $d^{(0)}_n$ against $\theta_n$]{
\includegraphics[width= .45\textwidth]{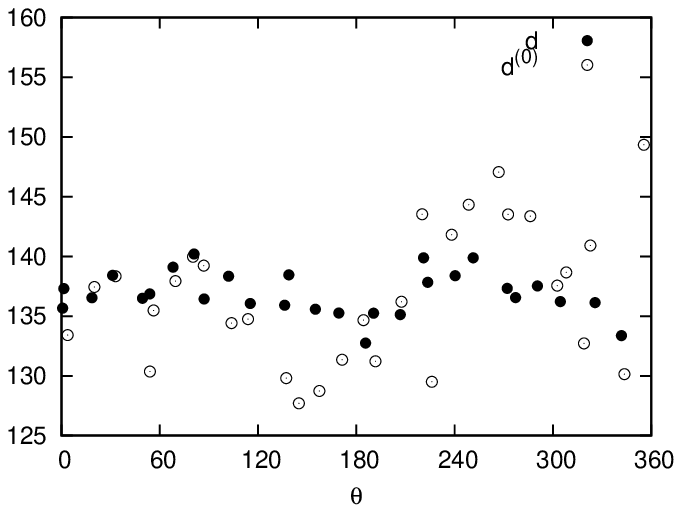}
\label{rutishauser:subfig3}
}
\subfigure[$\log r_n$ and $\log r^{(0)}_n$ against $n$]{
\includegraphics[width= .45\textwidth]{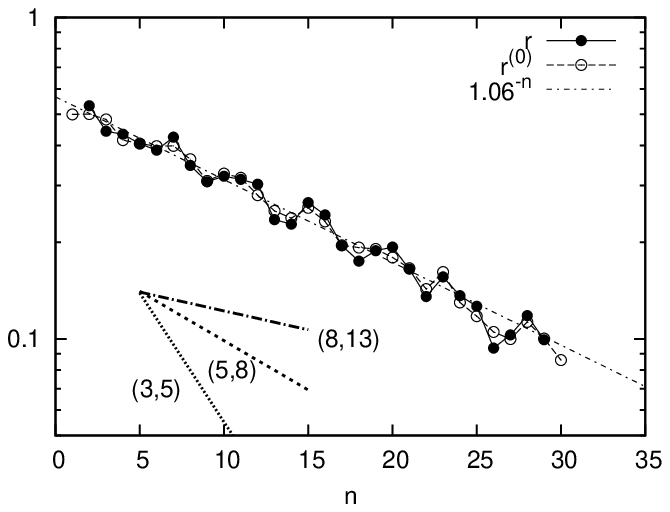}
\label{rutishauser:subfig4}
}
\caption[]{
Results for a normal phyllotaxis of young leaves of {\it Suaeda vera},
 whose pattern shown in  (a) is taken after Fig.~1a of \cite{rutishauser98}. 
Radiating lines in (a) represent a divergence diagram
(cf. the caption of Fig.~\ref{spiral} and Sec.~\ref{sec:floatingcenter}). 
\label{rutishauser}
}
\end{figure}

As a typical example of real systems,  
phyllotaxis of young leaves of {\it Suaeda vera} is 
shown in Fig.~\ref{rutishauser:subfig1} after Fig.~1a of \cite{rutishauser98}.  
The actual size of the pattern 
is about 1mm in diameter. 
Leaf position is optically read as marked in the original figure, 
which is a drawn figure based on a microtome section.  
A leaf is not a point. It has shape and size. 
In this case, and presumably in most past cases reported so far, 
position of a leaf is represented by the position of the main vascular
bundle when it is visually discernible. 
This paper does not take up 
cases in which position of leaves may not be specified unambiguously.  
%
%
A fixed center $O_0$ for $d^{(0)}$ in Figs.~\ref{rutishauser:subfig2} and \ref{rutishauser:subfig3} 
is set according to the original figure. 
Statistical results for divergence angle are $d^{(0)}=136.8 \pm 5.8^\circ$ by the fixed center $O_0$, 
$d=136.9 \pm 1.8^\circ $ by the floating center of divergence, 
and $d^{(5,8)}=137.0\pm 0.9^\circ$ by the floating center of parastichy. 
It is remarkable that no significant deviation
from the ideal limit angle of $137.5^\circ$ is found,  
although the divergence diagram  in Fig.~\ref{rutishauser:subfig1}
indicates that the pattern does not have a well-defined center. 
In the figures,  $d_n^{(5,8)}$ for $n=15$ is omitted because no solution
was found.

\cite{rutishauser98} has remarked angular width of leaf arc, $i$,  
as another quantity of interest. 
From the original figure, it is estimated as 
$i=66\pm 13^\circ$.
Oddly enough, the standard deviation of 
 the angular width of leaf arc is larger than that of divergence angle.  
If  the right tip of leaf arc is regarded as a representative point of the leaf, 
$d^{(0)}= 137.4 \pm 6.0^\circ$. 
If the left tip is used instead, $d^{(0)}= 136.1 \pm 5.8^\circ$. 
Thus, divergence angle does not depend on which part of leaf is regarded
as a representative point.
The results indicate that 
absolute positions of the leaves are not affected even though 
the angular width of each leaf arc fluctuates widely.  

%
%
%
%


%
Apparent lack of stability in nominal divergence angle $d^{(0)}$
(open circles in Fig.~\ref{rutishauser:subfig2}) is due
to systematic variations of the type expected for the first order
deformation (Fig.~\ref{rutishauser:subfig3}).  
The azimuth plot of $d$ in Fig.~\ref{rutishauser:subfig3} (closed circles) indicates 
 a slight indication of peaks at $\theta\simeq 80^\circ$ and
 $\theta\simeq 80+180^\circ$, signifying deformation of the second order.

A semi-log plot of $r_n$ and $r_n^{(0)}$ in Fig.~\ref{rutishauser:subfig4}
confirms the exponential rule (\ref{rn=an}). 
Straight lines labeled with $(3,5)$, $(5,8)$ and $(8,13)$
represent the exponential growth according to (\ref{loga2pisqrt5tau2n+1}) for $i=4,5$ and 6,  respectively.  
In the original figure, leaves are in contact with each other along 
 three contact parastichies 3, 5 and 8. 
The growth rate $\log a$ has to be estimated in order 
to evaluate the phyllotaxis index (P.I.) in (\ref{P.I.log}). 
This may pose a methodological problem. 
The radial growth $r_n$ of a real system is neither continuous nor monotonic in $n$. 
There are systematic variations in $r_n^{(0)}$ and $r_n$, 
particularly because the pattern does not have a definite center. 
Accordingly, 
a bad method or a bad choice of leaves
may give an unwanted result $\log a<0$ for some values of $n$. 
\cite{rutishauser98} has evaluated $a\simeq 1.06$ 
based on two ratios for three leaves, namely 2, 18 and 31 in
Fig.~\ref{rutishauser:subfig1} (30, 14 and 1 in the original figure). 
In the presence of systematic variations, 
a curve fitting method objectively gives a more accurate result.  
As shown in Fig.~\ref{rutishauser:subfig4}, 
the least-squares fitting method with the function $r_0 \exp(-n \log a)$ 
gives $\log a   = 0.0580\pm 0.0014$ ($a=1.0597\pm 0.0015$)
and $\log a   = 0.0593\pm 0.0023$ ($a=1.0611 \pm 0.0024$)
for $r_n^{(0)}$ and  $r_n$, respectively. 
Therefore,  P.I.=4.18 by (\ref{P.I.log}). 
The pattern is close to an orthogonal $(5,8)$ system with P.I.=4. 

\subsection{Effect of gibberellic acid on Xanthium}
\label{egaonxanthium}

\begin{figure}
\centering
\subfigure[$(x_n, y_n)$]{
\includegraphics[height= .3\textwidth]{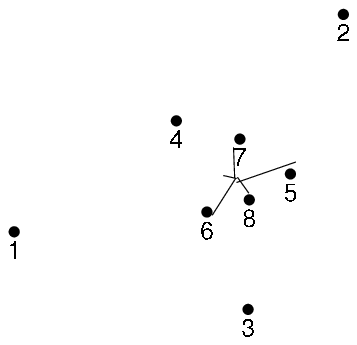}
\label{me2:subfig1}
}
\subfigure[$d_n$ and $d^{(0)}_n$ against $n$]{
\includegraphics[width= .45\textwidth]{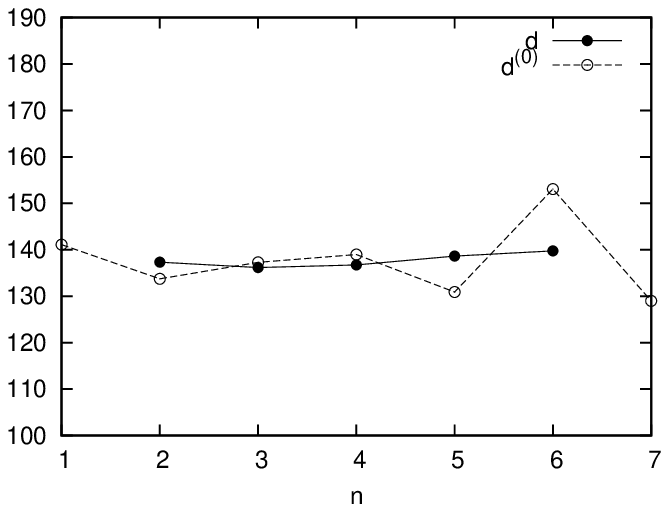}
\label{me2:subfig2}
}
\subfigure[$d_n$ and $d^{(0)}_n$ against $\theta_n$]{
\includegraphics[width= .45\textwidth]{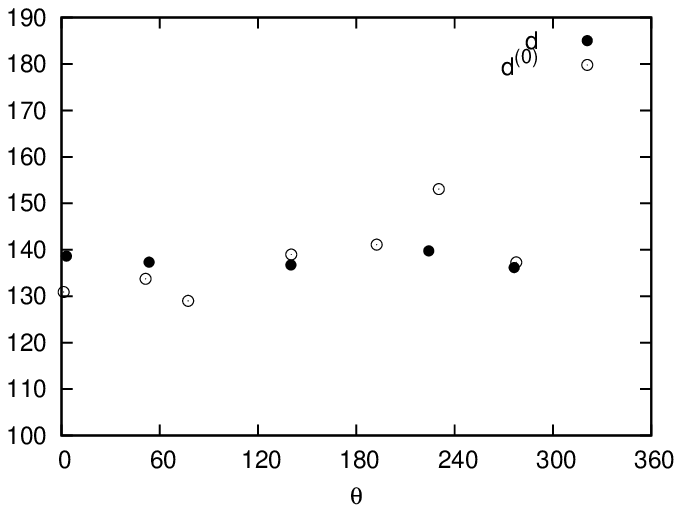}
\label{me2:subfig3}
}
\subfigure[$\log r_n$ and $\log r^{(0)}_n$ against $n$]{
\includegraphics[width= .45\textwidth]{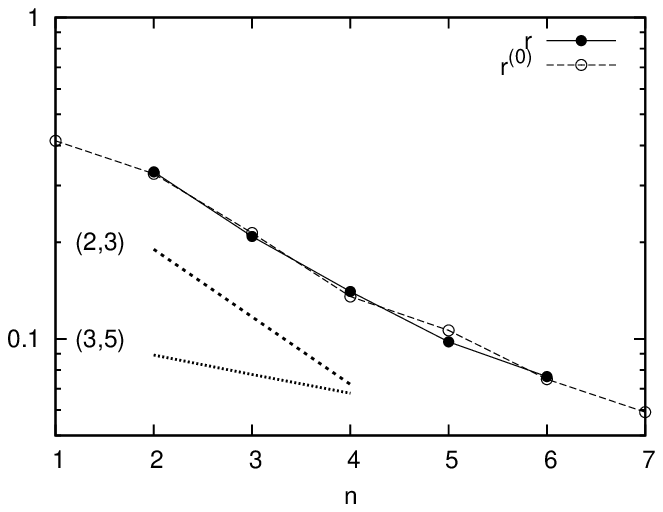}
\label{me2:subfig4}
}
\caption[]{
Results for a control shoot of {\it Xanthium}, 
after the bottom photo-micrograph in Fig.~6 of \cite{me77}. 
Compare with Fig.~\ref{me1}. 
\label{me2}
}
\end{figure}
\begin{figure}
\centering
\subfigure[$(x_n, y_n)$]{
\includegraphics[height= .3\textwidth]{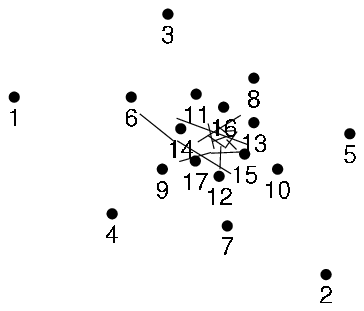}
\label{me1:subfig1}
}
\subfigure[$d_n$ and $d^{(0)}_n$ against $n$]{
\includegraphics[width= .45\textwidth]{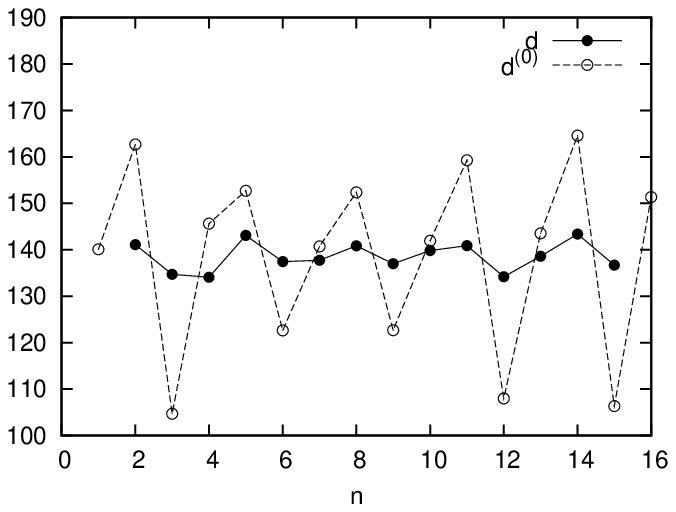}
\label{me1:subfig2}
}
\subfigure[$d_n$ and $d^{(0)}_n$ against $\theta_n$]{
\includegraphics[width= .45\textwidth]{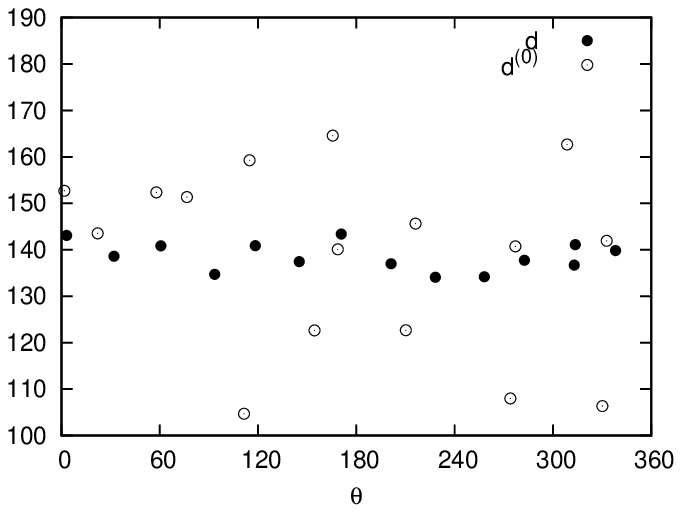}
\label{me1:subfig3}
}
\subfigure[$\log r_n$ and $\log r^{(0)}_n$ against $n$]{
\includegraphics[width= .45\textwidth]{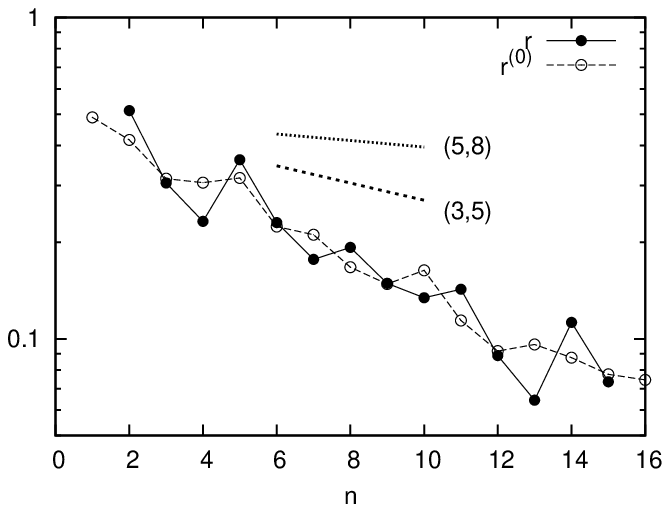}
\label{me1:subfig4}
}
\caption[]{
Results for a gibberellic acid treated shoot of {\it Xanthium}, 
after the top photo-micrograph in Fig.~6 of \cite{me77}. 
Compare with Fig.~\ref{me2}. 
\label{me1}
}
\end{figure}

Figs.~\ref{me2} and \ref{me1} 
are results obtained after the bottom and top photomicrographs in Fig.~6
of \cite{me77},  which are 
control and gibberellic acid (GA) treated vegetative shoots of {\it Xanthium}, respectively.  
The mean and standard deviation of divergence angle are 
$d^{(0)}=137.73\pm 8.03^\circ$ and $d=137.74\pm 1.46^\circ$ for the former, 
while $d^{(0)}=138.71\pm 19.93^\circ$ and $d=138.55\pm 3.09^\circ$ for the latter. 
The pattern of the control shoot in Fig.~\ref{me2:subfig1} indicates a well-defined center, 
while the treated shoot in Fig.~\ref{me1:subfig1} does not have a center.  
As a result, $d^{(0)}$ of the treated plant shows wild fluctuations. 
%
Thus, the GA treatment not only decreases the growth rate $\log a$
(the slope of Figs.~\ref{me2:subfig4} and \ref{me1:subfig4}), 
but destabilize divergence angle appreciably (Fig.~\ref{me1:subfig2}). 
Nonetheless,  the treatment does not affect 
the mean divergence angle. 
These results are consistent with analysis by \cite{me77} based on 
the exponential growth (\ref{rn=an}). 
The exponential growth is corroborated qualitatively, if not
quantitatively, as shown in Figs.~\ref{me2:subfig4} and \ref{me1:subfig4}, 
where reference lines labeled with parastichy pairs of Fibonacci numbers $(F_i, F_{i+1})$
represent the exponential growth with the constant rate 
according to (\ref{loga2pisqrt5tau2n+1}). 

%


\subsection{Helianthus tuberosus}
\label{subsec:ht}

\subsubsection{Lucas system: (11,18)}

%

\begin{figure}
\centering
\includegraphics[width= .45\textwidth]{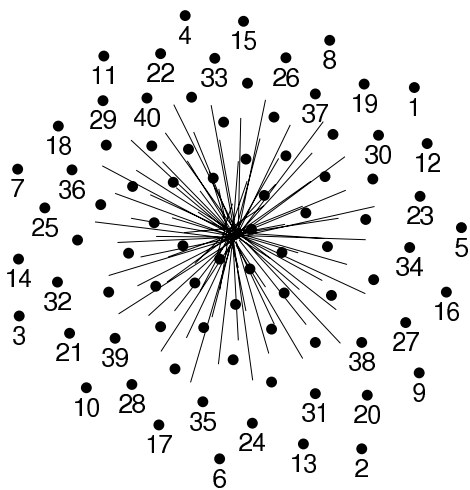}
\caption[]{
Lucas phyllotaxis on a capitulum of {\it Helianthus tuberosus} 
after Fig.~2 of \cite{rrb91}. 
\label{spiralrrb2}
}
\end{figure}
\begin{figure}
\centering
\subfigure[$d_n$ and $d^{(0)}_n$ against $n$]{
\includegraphics[width= .45\textwidth]{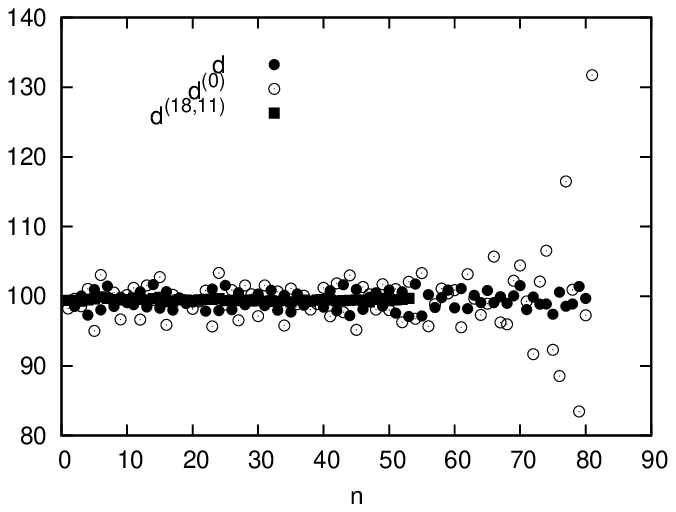}
\label{rrb2:subfig2}
}
\subfigure[$d_n$ and $d^{(0)}_n$ against $\theta_n$]{
\includegraphics[width= .45\textwidth]{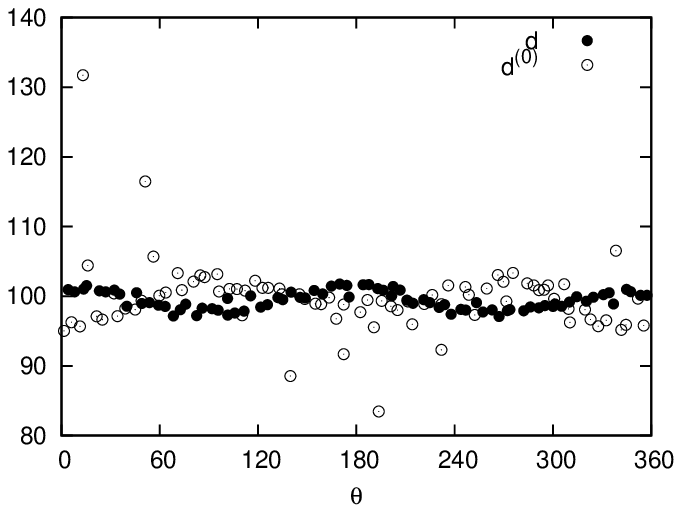}
\label{rrb2:subfig3}
}
\subfigure[$\log r_n$ and $\log r^{(0)}_n$ against $n$]{
\includegraphics[width= .45\textwidth]{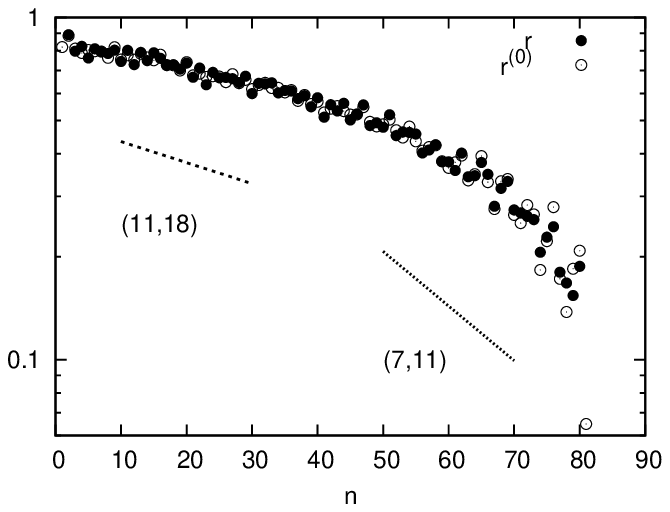}
\label{rrb2:subfig4}
}
\subfigure[$r_n$ and $r^{(0)}_n$ against $n$]{
\includegraphics[width= .45\textwidth]{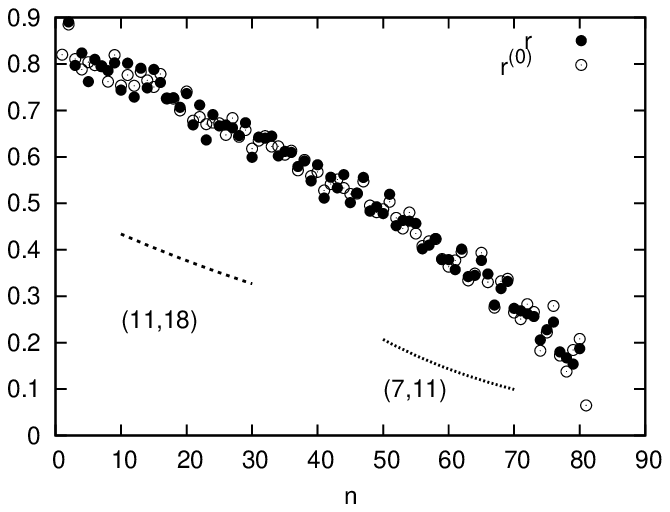}
\label{rrb2:subfig5}
}
\caption[]{
Results for a capitulum of {\it Helianthus tuberosus} plotted in Fig.~\ref{spiralrrb2}. 
(After Fig.~2 of \cite{rrb91}). 
\label{rrb2}
}
\end{figure}

In Figs.~\ref{spiralrrb2} and \ref{rrb2}, 
results are shown for a capitulum of {\it Helianthus tuberosus} after Fig.~2 of \cite{rrb91}. 
The size of the pattern is about 13mm in diameter. 
Here and below, digitized position of florets is taken after \cite{rrb91}, 
although the numberings are different from the original figures. 
This is a Lucas 
system with the limit divergence angle of 
\begin{equation}
 \alpha_0=\frac{1}{3+\tau^{-1}}, 
\label{alpha0=frac1/3+tau-1} 
\end{equation}
or $d=2\pi \alpha_0=99.5^\circ$. 
A nominal center $O_0$ is set at the center of gravity, or by 
\begin{equation}
\sum_n \overrightarrow{O_0P_n} =\vec{0}.
\label{sumn=Oon=vec0}
\end{equation}
The center of gravity is a good 
choice for the fixed center
especially for a system comprising a large number of pattern units. 
Results for divergence angle are 
$d^{(0)}=99.72\pm 5.35^\circ$, $d=99.42 \pm 1.28^\circ$
and $d^{(18,11)}=99.47\pm 0.15^\circ$. 
Remark the accuracy attained in $d^{(18,11)}$. 
Thus, the center of parastichy comes into its own in a high phyllotaxis pattern. 
In a closely packed pattern, 
young seeds in a central region are amenable to irregular displacements 
caused by inward pressure due to old seeds in an outer region. 
This is observed as wide fluctuations of $d^{(0)}$ for $n>70$ in Fig.~\ref{rrb2:subfig2}.   
As shown below, the secondary displacement 
appears to be a primary source of variations in divergence angle of a packed pattern. 
The divergence angles in Fig.~\ref{rrb2:subfig3} indicate systematic
variations of the compression type.  
The amplitude $\Delta d^{(0)}\simeq 4^\circ$ implies $C_2\simeq 0.07$ by
(\ref{Deltadsimeq565}), 
although the deformation is apparently indiscernible from Fig.~\ref{spiralrrb2}. 
In general, deviations of $r_n$ from an exponential dependence 
give rise to shifts in parastichy numbers. 
In Figs.~\ref{rrb2:subfig4} and \ref{rrb2:subfig5}, 
curves labeled with $(11, 18)$ and $(7, 11)$
represent the exponential dependence (\ref{rn=an}) with (\ref{loga=2pisqrt}) 
for $\frac{p}{q}=\frac{5}{18}$,  $\frac{p'}{q'}=\frac{3}{11}$
and $\frac{p}{q}=\frac{2}{7}$,  $\frac{p'}{q'}=\frac{3}{11}$,
respectively.
The limit divergence angle of (\ref{alpha0=frac1/3+tau-1}) is used for
$\alpha_0$.  
The $n$-dependence of $r_n$ is neither exponential (\ref{rn=an}),  nor
square root (\ref{rn=Asqrtn}). 
It is rather close to linear (Fig.~\ref{rrb2:subfig5}). 
Accordingly, conspicuous parastichies change continuously 
from $(11,18)$ near the rim to $(7,11)$ near the center. 
The apparent up shift of the parastichy pair caused by 
gradual decrease of $\log a$ is called {\it rising phyllotaxis}.  
The pattern in Fig.~\ref{spiralrrb2}
shows {\it falling phyllotaxis}, the downshift of the parastichy pair 
from $(11,18)$ to $(7,11)$, and even down to $(4,7)$. 
It should be remarked that 
new cell primordia arise from the rim towards the center, 
 not vice versa as often assumed wrongly in theoretical models (cf. (\ref{rn=Asqrtn})). 
Mathematically, the shift of parastichy corresponds to 
the fact that the function $\log r_n$ is convex upward as shown in Fig.~\ref{rrb2:subfig4}. 
Despite the gradual shift in parastichy numbers, 
divergence angle is not affected (Fig.~\ref{rrb2:subfig2}). 
%
%
If phyllotaxis is to be judged by the divergence angle, 
there is no sign of change in phyllotaxis.  




\subsubsection{Trijugate system: 3 (3,5)}
\label{sec:trijugate}

\begin{figure}[t]
\centering
\includegraphics[width= .7\textwidth]{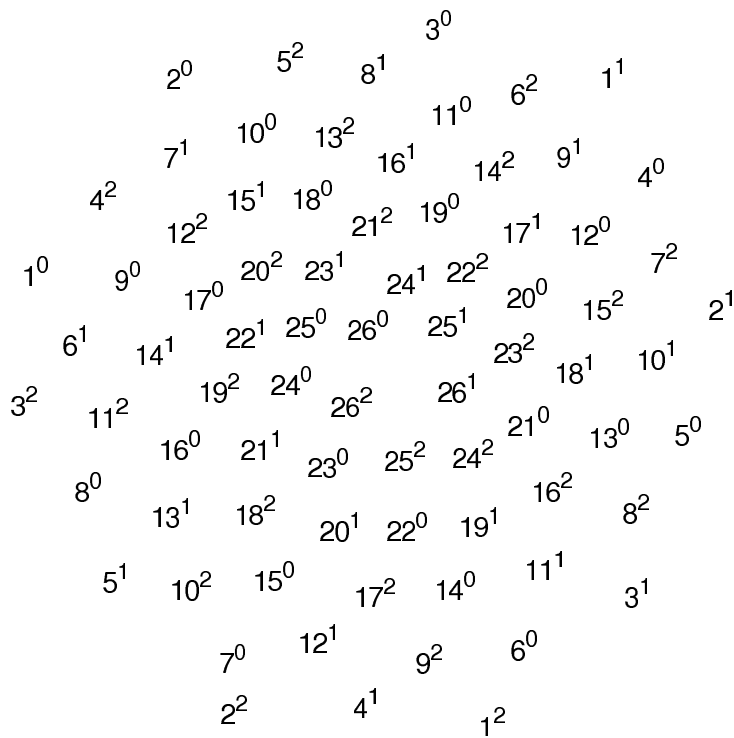}
\caption[]{
A rare trijugate $3(3,5)$ phyllotaxis 
on a sectioned capitulum of  {\it Helianthus tuberosus}.  
Adapted after Fig.~3 of \cite{rrb91}. 
Seeds are indexed as $n^j$ 
according to the numbering system $(n,j)$ explained 
in Sec.~\ref{sec:whorled}. 
\label{spiralrrb3}
}
\end{figure}
\begin{figure}
\centering
\subfigure[]{
\includegraphics[width= .45\textwidth]{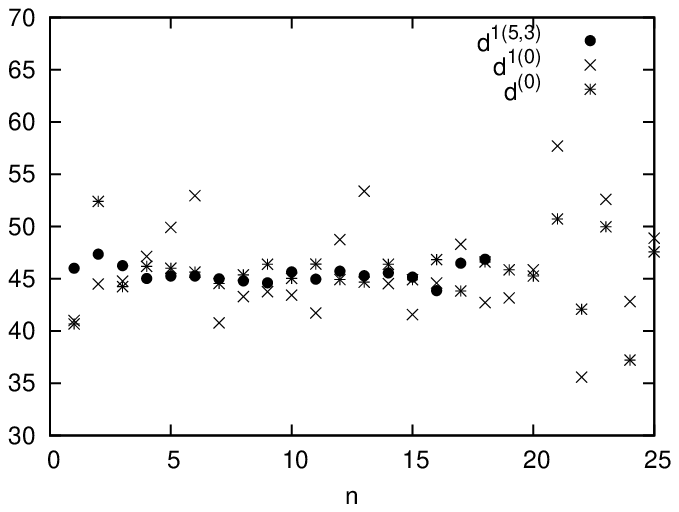}
\label{rrb3:subfig1}
}
\subfigure[]{
\includegraphics[width= .45\textwidth]{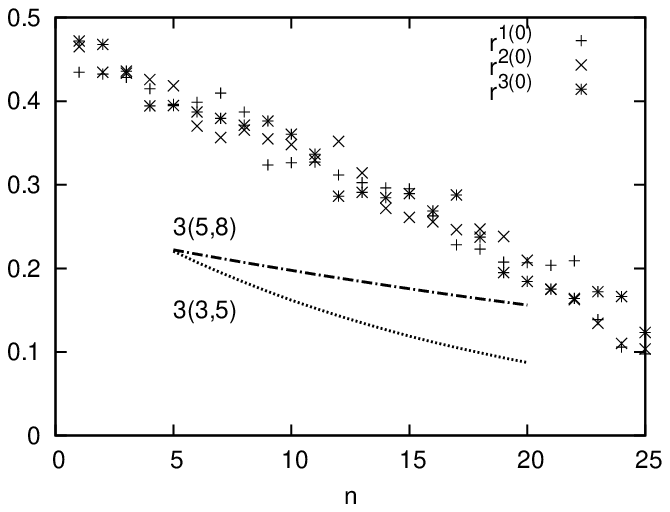}
\label{rrb3:subfig2}
}
\caption{
Results for a capitulum of {\it Helianthus tuberosus} 
plotted in Fig.~\ref{spiralrrb3} (after Fig.~3 of \cite{rrb91}). 
(a) $d_n^{1(5,3)}$, $d_n^{1(0)}$ and 
$d_n^{(0)}=(d_n^{1(0)}+d_n^{2(0)}+d_n^{3(0)})/3$, and 
(b) $r_n^{1(0)}$, $r_n^{2(0)}$ and $r_n^{3(0)}$
are plotted against the leaf index $n$. 
They are evaluated in terms of the nominal center $O_0$ set by (\ref{sumn=Oon=vec0}). 
\label{rrb3}
}
\end{figure}

Figs.~\ref{spiralrrb3} and \ref{rrb3} are results for 
a rare trijugate pattern $(J=3)$ of {\it Helianthus tuberosus} 
based on Fig.~3 of \cite{rrb91}. 
The size of the pattern is about 8mm in diameter. 
In terms of the nominal center $O_0$ set by (\ref{sumn=Oon=vec0}),  one obtains  
 $3d^{j(0)}=137.4\pm 14.7^\circ$, 
$137.7\pm 12.9^\circ$ and $135.3\pm 19.2^\circ$ for $j=1,2$ and 3, respectively,  
where 
\[
 d_n^{j(0)} = \angle P_n^j O_0 P_{n+1}^j 
\]
is divergence angle depending on the jugacy index $j$. Note that $j=3$ is equivalent to $j=0$. 
In Fig.~\ref{spiralrrb3}, the seed point $P_n^j$ is denoted as $n^j$. 
For every integer $n$, 
the trijugate pattern has three organs of approximately 120~degrees
apart from each other, 
which are distinguished by the jugacy index $j$. 
When averaged over $j$,  nominal divergence angle is 
$3d^{(0)}=136.8 \pm 9.0^\circ$, 
showing a large standard deviation. 
As shown by $d_n^{1(0)}$ in Fig.~\ref{rrb3:subfig1}, 
variations in divergence angle of the multijugate system 
are substantially larger than that of a simple spiral system. 

The radial coordinate 
\[
 r_n^{j(0)} = |\overrightarrow{O_0 P_n^j}|, 
\]
depends linearly on $n$ with less remarkable fluctuations. 
In Fig.~\ref{rrb3:subfig2}, 
curves labeled with $3(3,5)$ and $3(5,8)$ are drawn by (\ref{Jloga2pisqrt5tau2n+1}) 
for $J=3, i=4$ and $J=3, i=5$, respectively. 
The results suggest that fluctuations are due to displacements
in the angular direction. 
Indeed, 
large fluctuations 
$\Delta^{(0)}=0 \pm 6.85^\circ$ are observed for (\ref{Deltanj=thetanj+1}). 
Multijugate systems are characterized by this large angular
fluctuations. 
For this reason, 
 the method of the center of divergence is barely applicable. 
By means of the parastichy center in Sec.~\ref{sec:whorled}, 
we get $3d^{j(3,5)}=136.53\pm 2.55^\circ$, 
$137.85\pm 2.88^\circ$ and  
$137.43\pm 3.78^\circ$ for $j=1,2$ and 3, respectively. 
In total, 
$3d^{(3,5)}=137.28\pm 3.09^\circ$. 
The standard deviations are significantly reduced 
but still an order of magnitude larger than the spiral system of the last subsection. 
The absence of a center of rotation in a real $J$-jugate system 
is manifested by the very fact that 
expected rotational symmetry of order $J$ is badly approximate. 
This may not be apparent at a glance of Fig.~\ref{spiralrrb3}, 
but it is immediately checked if Fig.~\ref{spiralrrb3} is overlapped with
itself after being rotated by 120~degrees,  
or, e.g., by comparing three pairs of 
$(2^0, 7^{1})$, $(2^1, 7^{2})$ and $(2^2, 7^{0})$. 
The rotated pattern does not coincide with the original one.

%
%

\subsection{Helianthus annuus}
\label{subsec:ha}

\begin{figure}[t]
\centering
\includegraphics[width= .7\textwidth]{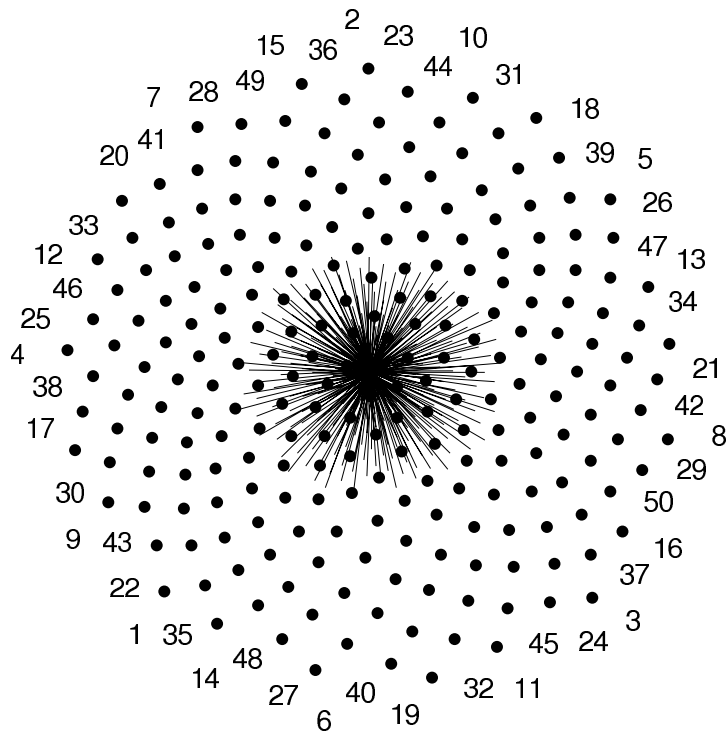}
\caption[]{
Developing florets on a capitulum of {\it Helianthus annuus} 
after Fig.~4 of \cite{rrb91}. 
\label{spiralrrb4}
}
\end{figure}
\begin{figure}
\centering
\subfigure[$d_n$ and $d^{(0)}_n$ against $n$]{
\includegraphics[width= .35\textwidth]{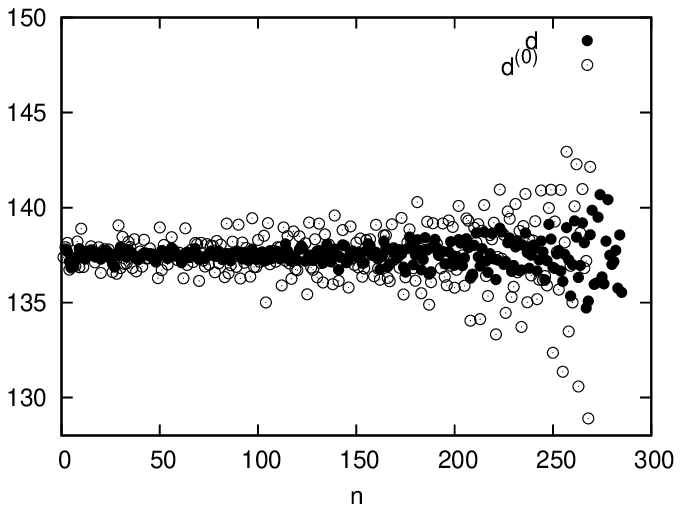}
\label{rrb4:subfig2}
}
\subfigure[$d_n$ and $d^{(0)}_n$ against $\theta_n$]{
\includegraphics[width= .35\textwidth]{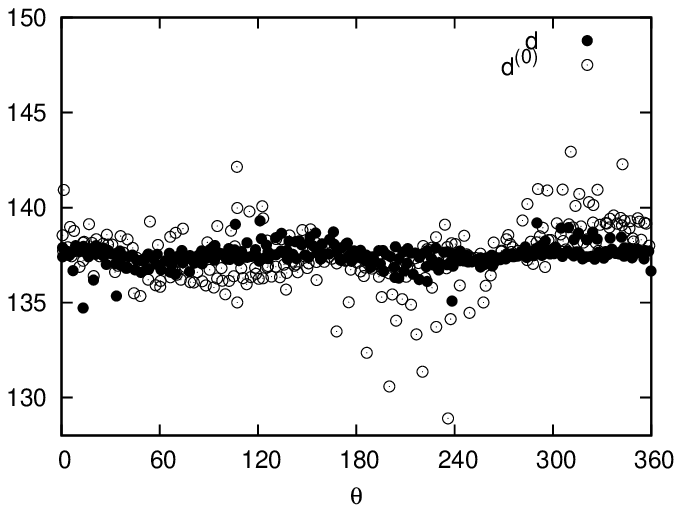}
\label{rrb4:subfig3}
}
\subfigure[$\log r_n$ and $\log r^{(0)}_n$ against $n$]{
\includegraphics[width= .35\textwidth]{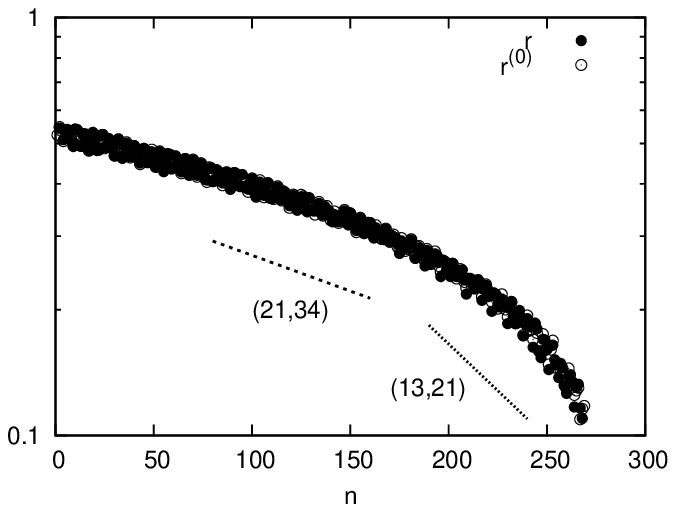}
\label{rrb4:subfig4}
}
\subfigure[$r_n$ and $r^{(0)}_n$ against $n$]{
\includegraphics[width= .35\textwidth]{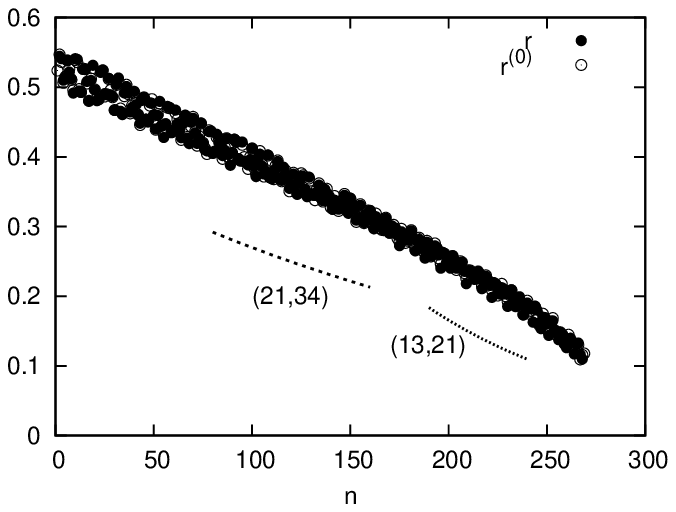}
\label{rrb4:subfig5}
}
\subfigure[$r_n^2$ against $n$]{
\includegraphics[width= .35\textwidth]{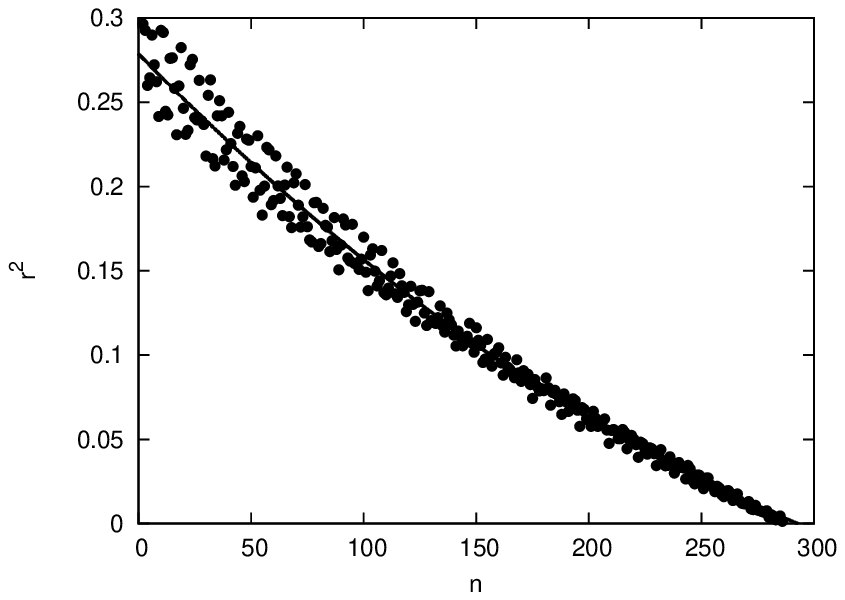}
\label{rrb4:subfig6}
}
\subfigure[$d_n$ against the phyllotaxis index]{
\includegraphics[width= .35\textwidth]{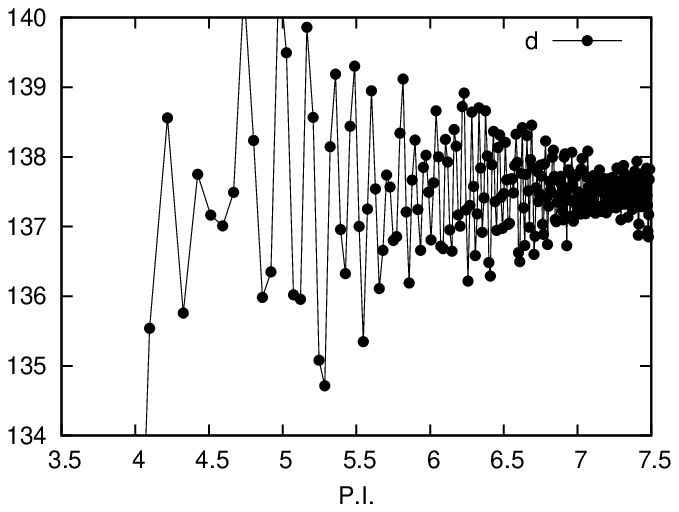}
\label{rrb4:subfig7}
}
\subfigure[$d_n$ against $\theta_n$]{
\includegraphics[width= .35\textwidth]{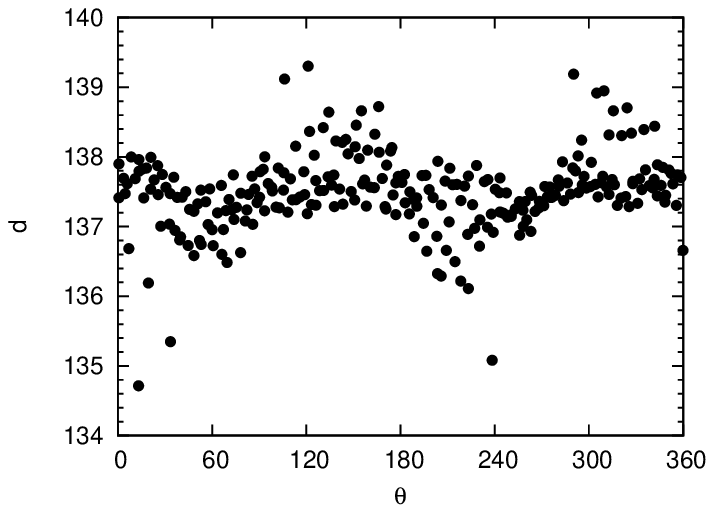}
\label{rrb4:subfig8}
}
\subfigure[$\Delta r_n/r_n$ against $\theta_n$]{
\includegraphics[width= .35\textwidth]{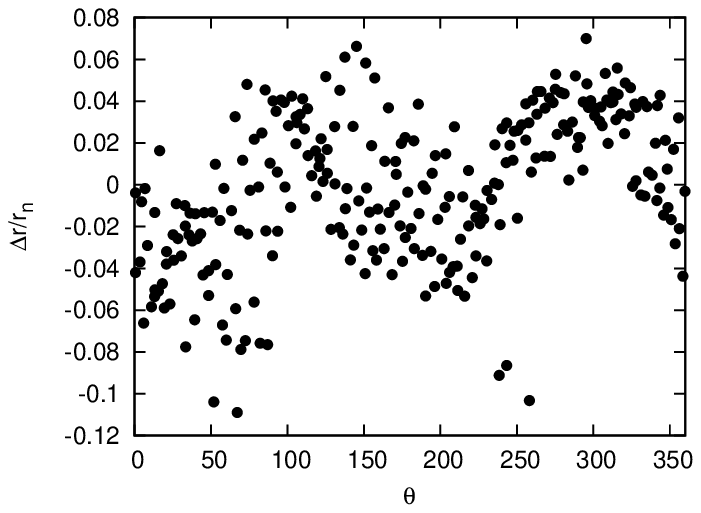}
\label{rrb4:subfig9}
}
\caption[]{
Results for a capitulum of {\it Helianthus annuus} in Fig.~\ref{spiralrrb4}
(after Fig.~4 of \cite{rrb91}). 
\label{rrb4}
}
\end{figure}

\begin{figure}[t]
\centering
\includegraphics[width= 0.9\textwidth]{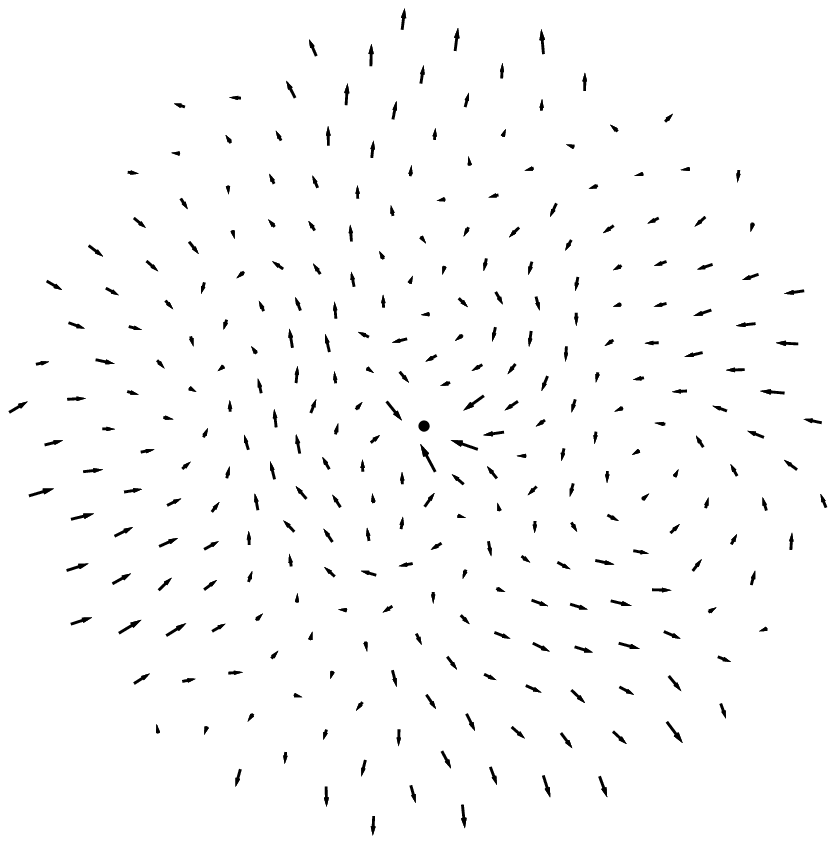}
\caption[]{
Displacement from regular position of 
florets on the capitulum of {\it Helianthus annuus} in
 Fig.~\ref{spiralrrb4} (cf. Fig.~\ref{rrb4}). 
After Fig.~4 of \cite{rrb91}.
\label{rrb4vector}
}
\end{figure}


Figs.~\ref{spiralrrb4} and \ref{rrb4} are for developing florets on a
capitulum of {\it Helianthus annuus} after Fig.~4 of \cite{rrb91}. 
The actual size is about 9 cm in diameter. 
Numerical results for divergence angle 
are $d^{(0)}=137.49\pm 1.64^\circ$, $d=137.49\pm 0.56^\circ$
and $d^{(34,21)}=137.50\pm 0.07^\circ$. 
In numerical calculation, 
some inner seeds fail to give the solution for the center of parastichy,  
owing to significant displacement from regular position.  
\cite{rrb91} have noted large fluctuations of $d^{(0)}$, 
open circles in Fig.~\ref{rrb4:subfig2}. 
The present study reveals that 
the fluctuations are not random but mainly systematic
variations due to local deformation of the pattern, 
as indicated by the azimuth plots in Figs.~\ref{rrb4:subfig3} and
\ref{rrb4:subfig8}. 

The radius in Fig.~\ref{rrb4:subfig6} is fitted with 
\begin{equation}
 r_n= r_0 \sqrt{1-bn+cn^2} 
\label{rn=r0sqrt1-bn+cn2}
\end{equation}
to give $r_0= 0.528\pm 0.002$ (arbitrary), 
$b = (4.88\pm 0.07)\times 10^{-3}$, and 
$c=  (5.01\pm 0.27)\times 10^{-6}$.
Thus, the area per seed,  
\[
\frac{{\rm d}}{{\rm d}(-n)} (\pi r_n^2) =\pi r_0^2 (b + 2c(-n)), 
\]
depends on the seed index $n$.
The positive value of the coefficient $c$ signifies that inner seeds are smaller than
outer seeds, 
as expected physically. 
The relative change in size is ${2c}/{b} = 0.002$, namely 0.2 percent per seed.  
Accordingly,  the innermost seed for $n=287$ is estimated to be about half 
the size of the seeds around the rim.
This is indeed seen from the original figure.
Scatterings of divergence angles of inner seeds (Fig.~\ref{rrb4:subfig2}) 
suggest that the physical effect to reduce the size of seeds affects their position and divergence angles accordingly. 
In terms of the continuous monotonic function (\ref{rn=r0sqrt1-bn+cn2}), 
the $n$-dependent plastochron ratio is evaluated as 
\begin{equation}
\log a=- \frac{{\rm d}}{{\rm d}n}\log r_n 
=\frac{b-2cn}{2({1-bn+cn^2})}, 
\end{equation}
with which the phyllotaxis index is calculated by (\ref{P.I.log}). 
In Fig.~\ref{rrb4:subfig7}, 
the divergence angle $d_n$ is shown 
against the phyllotaxis index.
Note that $\rm P.I.=5, 6$ and 7 represent
$(8,13)$, $(13,21)$ and $(21,34)$ orthogonal systems, respectively. 
As the capitulum develops, 
the phyllotaxis index decreases
and variance in $d$ increases accordingly (falling phyllotaxis). 
However, the mean value of $d$ is constantly held at the limit divergence angle. 
The deviation $\Delta r_n\equiv  r_n- r_0 \sqrt{1-bn+cn^2}$ 
normalized by $r_n$ is plotted against the azimuthal angle $\theta$ in
Fig.~\ref{rrb4:subfig9}. 
Variations in $r_n$ show a similar dependence of $\theta$ as $d_n$ in Fig.~\ref{rrb4:subfig8}, 
namely the dependence by deformation of the second order
(Sec.~\ref{sec:deformation}). 
Amplitudes of variations in 
Figs.~\ref{rrb4:subfig8} and \ref{rrb4:subfig9} are compared with 
Figs.~\ref{sampleazim} and \ref{sampleazimdelr} for $C_2= 0.05$. 
Displacements of seeds are explicitly shown in Fig.~\ref{rrb4vector}, where 
each vector represents shift in position of a seed 
from its regular position determined by the mean divergence angle,  
virtually equal to the limit divergence angle. 
The figure indicates that (i) the pattern is compressed horizontally
(stretched vertically)
 and (ii) inner seeds are displaced appreciably by inward pressure. 
These are consistent with the above observations, 
namely systematic modulation of divergence angle  and decrease in size of inner seeds. 
The latter is also consistent with the observation that 
seeds near the center are even aborted (\cite{rrb91}). 


(\ref{rn=r0sqrt1-bn+cn2}) is transformed to  
\begin{equation}
 r_n= r_0 \sqrt{(n_c-n)
\left(1+ \frac{n_c-n}{n_0}
\right)
}. 
\label{rn=r0sqrtnc-n}
\end{equation}
By definition, 
an integer $n_c$ is an index number of the innermost seed at the center, $r_{n_c}=0$. 
In practice, however, 
the index $n_c$ of the last seed may not be determined 
without ambiguity, 
for it depends on 
position of a nominal center and whether or not to count aborted seeds. 
Be that as it may, 
the numerical order of the index system
 can be reversed 
by formally replacing $n_c-n$ with $n$. 
Consequently, (\ref{rn=r0sqrtnc-n}) is equivalent to 
\begin{equation}
 r_n=r_0
\sqrt{n 
 \left(
1+ \frac{n}{n_0}
\right)
}. 
\label{rn=r0sqrtn1+n}
\end{equation}
in the reversed index system. 
Fig.~\ref{rrb4:subfig6} shows that the radial component $r_n$ is 
better fitted with (\ref{rn=r0sqrtn1+n})
than the square-root dependence of (\ref{rn=Asqrtn}). 
\cite{rrb91} have investigated 
products of algebraic,  logarithmic, and exponential dependences of $r_n$. 
The functional form of (\ref{rn=r0sqrtn1+n})
 has not been considered previously, 
although it has wider practical applicability. 
A new constant ${n_0}$ in (\ref{rn=r0sqrtn1+n}) roughly signals 
the index $n$ of a seed for which the square-root dependence
(\ref{rn=Asqrtn}) begins to fail; 
$r_n\propto \sqrt{n}$ for $n\ll n_0$ whereas $r_n\propto n$ for  $n\gg n_0$. 
As inner seeds $n\sim 0$ are likely to be displaced significantly, 
the limit exponent $\epsilon$  $(\simeq 0.5)$ of the power dependence near the center 
$r_n\propto n^{\epsilon}$ ($n\rightarrow 0$) would be of little physical significance. 
As a rough estimate, 
we get $n_0\sim 800$ for this sample,
while $n_0\sim 20$ for Fig.~\ref{rrb3:subfig2}.

\subsection{Artichoke}

\begin{figure}
\centering
\includegraphics[width= .7\textwidth]{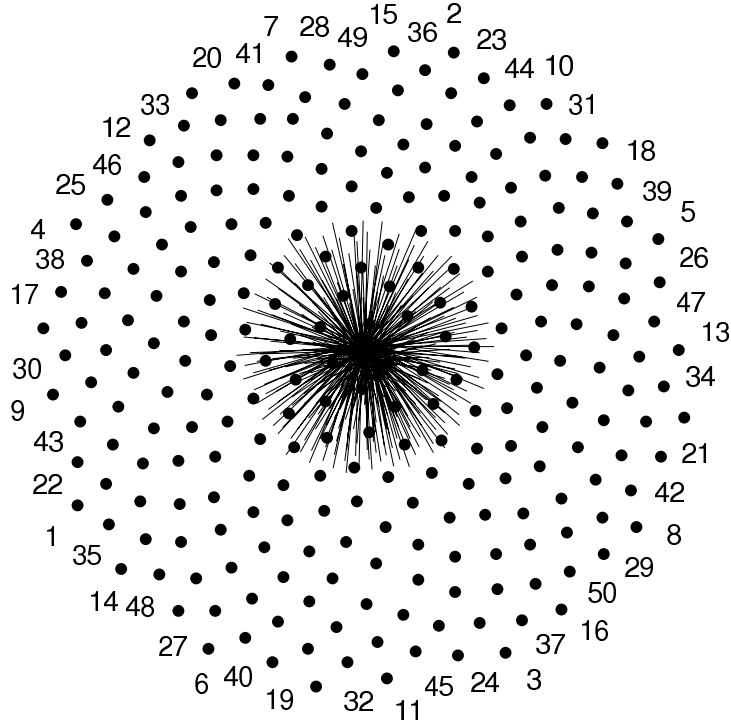}
\caption[]{
An artichoke capitulum after Fig.~9A of \cite{hjwzagd06}. 
\label{hottonspiral}
}
\end{figure}
\begin{figure}
\centering
\subfigure[$d_n$ and $d^{(0)}_n$ against $n$]{
\includegraphics[width= .35\textwidth]{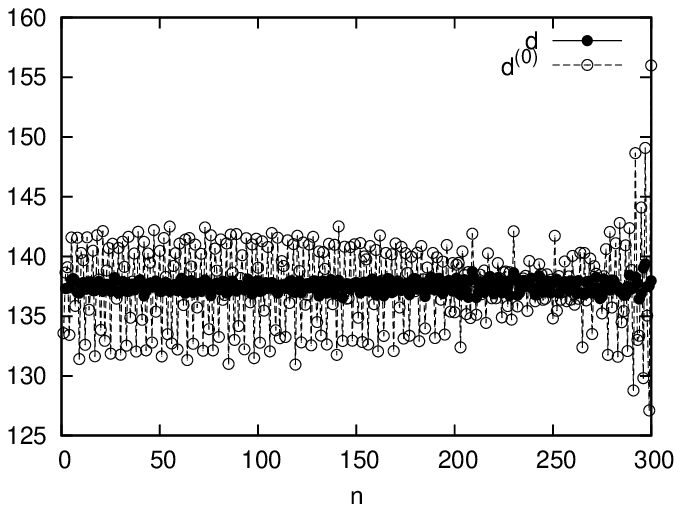}
\label{hotton:subfig2}
}
\subfigure[$d_n$ and $d^{(0)}_n$ against $\theta_n$]{
\includegraphics[width= .35\textwidth]{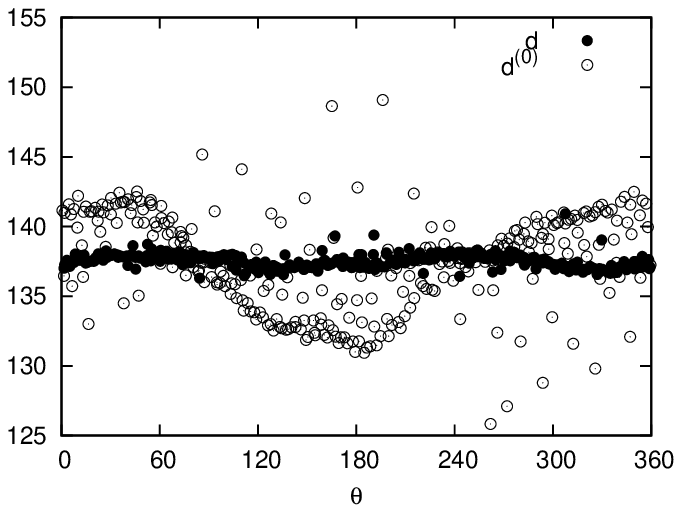}
\label{hotton:subfig3}
}
\subfigure[$\log r_n$ and $\log r^{(0)}_n$ against $n$]{
\includegraphics[width= .35\textwidth]{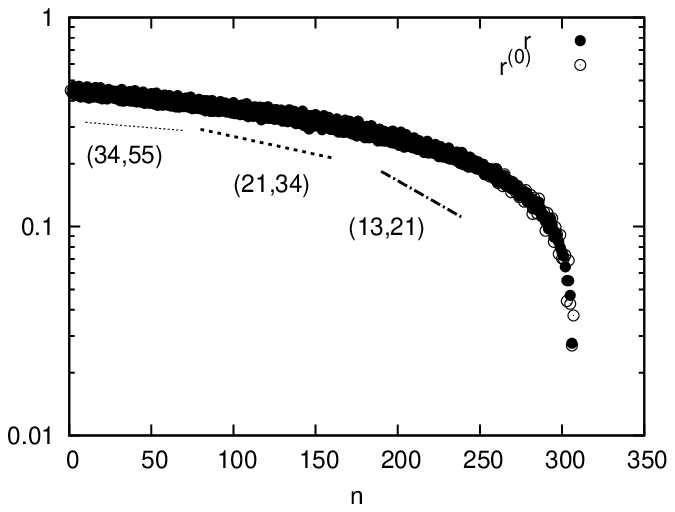}
\label{hotton:subfig4}
}
\subfigure[$r_n$ and $r^{(0)}_n$ against $n$]{
\includegraphics[width= .35\textwidth]{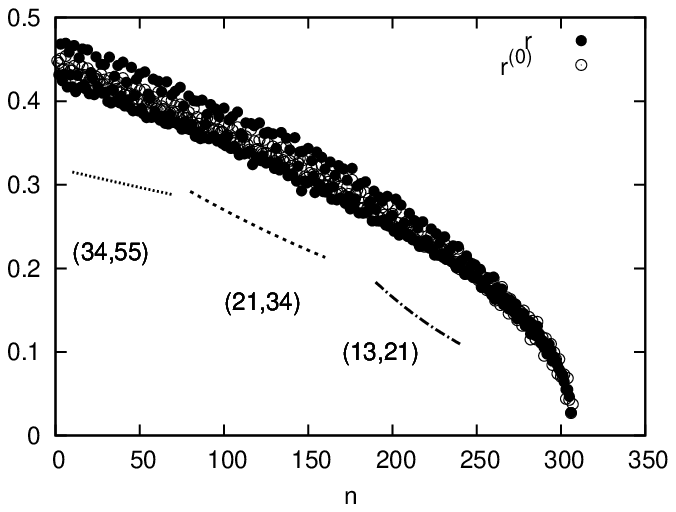}
\label{hotton:subfig5}
}
\subfigure[$d_n$ against $\theta$]{
\includegraphics[width= .35\textwidth]{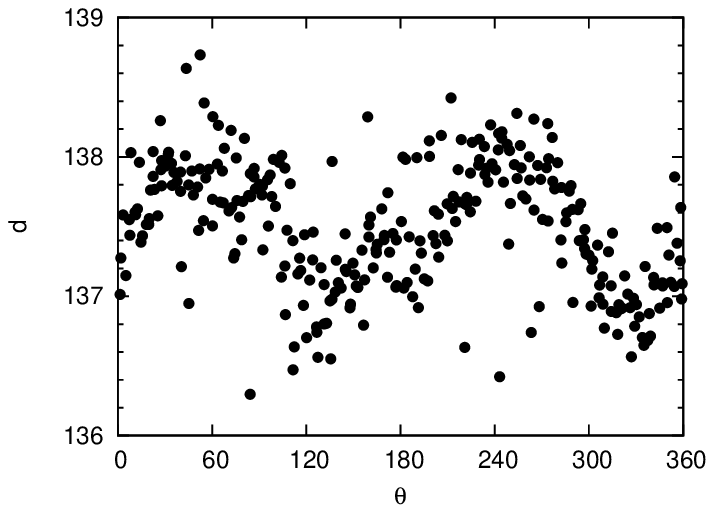}
\label{hotton:subfig6}
}
\subfigure[$r_n^2$ against $n$]{
\includegraphics[width= .35\textwidth]{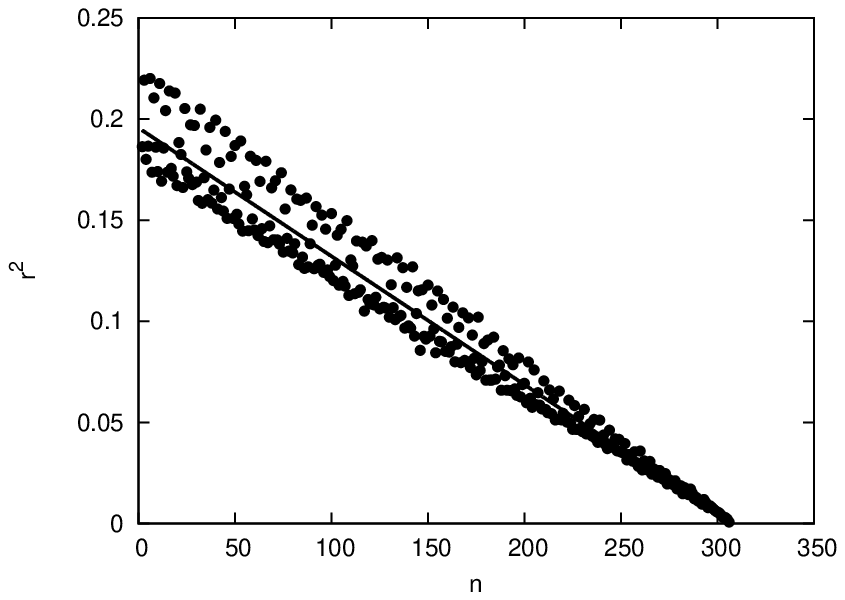}
\label{hotton:subfig7}
}
\subfigure[$\Delta r_n/r_n$ and $\Delta r_n^{(0)}/r_n^{(0)}$ against $\theta_n$]{
\includegraphics[width= .35\textwidth]{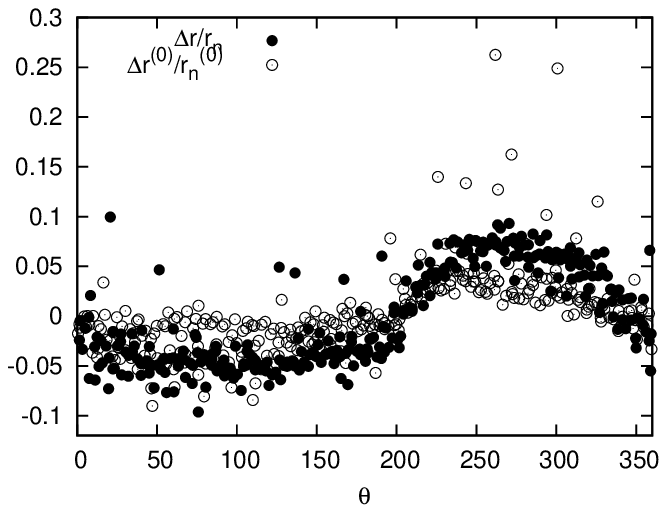}
\label{hotton:subfig8}
}
\subfigure[Displacement from regular position]{
\includegraphics[width= .35\textwidth]{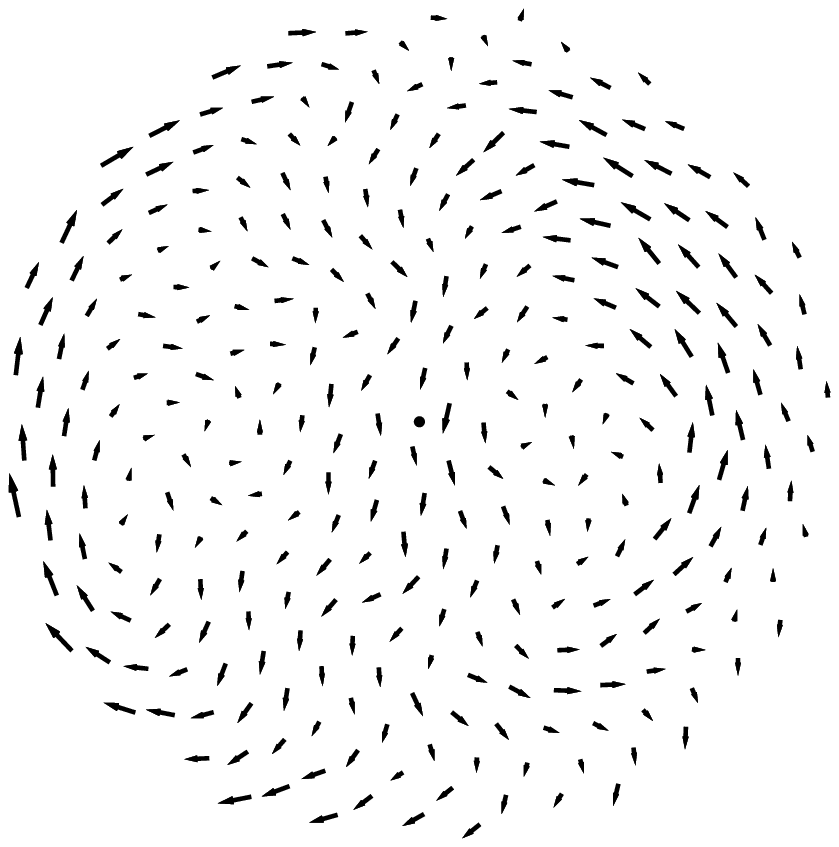}
\label{hotton:subfig9}
}
\caption[]{
Results for the artichoke capitulum in Fig.~\ref{hottonspiral}
(after Fig.~9A of \cite{hjwzagd06}).  
\label{hotton}
}
\end{figure}


Figs.~\ref{hottonspiral} and \ref{hotton} are results for an artichoke capitulum
based on Fig.~9A of \cite{hjwzagd06}.  
The size is about 4mm in diameter. 
Fig.~\ref{hottonspiral} closely resembles Fig.~\ref{spiralrrb4}, 
although the sizes differ by more than an order of magnitude.  
Position of primordia, which are round shaped and closely packed,  
 is read from the original figure with sufficient accuracy.
%
%
Numerical results obtained are $d^{(0)}=137.35\pm 5.03^\circ$, $d=137.53\pm 0.52^\circ$
and $d^{(34,21)}=137.50\pm 0.08^\circ$, 
all falling onto the limit divergence angle of (\ref{alpha0=tau-2}). 
The variations in $d^{(0)}$ are substantially
suppressed in $d$ and $d^{(34,21)}$ based on the floating centers. 
The standard deviation of $d^{(34,21)}$ is particularly noteworthy. 
The nominal center $O_0$ for $d^{(0)}$
is determined so as to minimize the standard deviation of $d^{(0)}$. 
Still, $d^{(0)}$ shows relatively large variations. 
If the center of gravity is chosen for $O_0$, 
one obtains $d^{(0)}=137.39\pm 5.03^\circ$ instead.  
Therefore, large deviations in $d^{(0)}$ are not due to misplacement of
a fixed center.
%
As a matter of fact, 
Fig.~\ref{hotton:subfig3} clearly indicates that 
deviations in divergence angle are due to systematic variations.  
As a rule, apparently wild variations of divergence angle 
that cannot be suppressed by a judicious choice
of a center should be suspected as a sign of no definite center by deformation. 
%
%
 Fig.~\ref{hotton:subfig6} reveals 
an unexpected result that 
remnant variations in $d_n$ are primarily not of random origin but 
of the compression type with modulation of $\Delta d\simeq 1^\circ$. 
%
%
As shown in Fig.~\ref{hotton:subfig7},  the radius $r_n$ shows 
a square-root dependence of (\ref{rn=Asqrtn}), 
namely $r_n= r_0 \sqrt{1-bn}$. 
The radius $r_n^{(0)}$, unlike the angle $d^{(0)}$, is generally
insensitive to the choice of a nominal center $O_0$, 
so that $r_n^{(0)}$ is not affected severely even 
if the nominal center $O_0$ is chosen inappropriately.
In Fig.~\ref{hotton:subfig8}, 
the deviation from the fit $\Delta r_n\equiv  r_n- r_0 \sqrt{1-bn}$
is plotted against the azimuthal angle $\theta$.
The systematic variations in $r_n$ have a different $\theta$-dependence from 
that of $d_n$ in Fig.~\ref{hotton:subfig6}. 
This is inexplicable by 
uniform deformation considered  in Sec.~\ref{sec:deformation}. 
As shown in Fig.~\ref{hotton:subfig9}, 
displacement of primordia from their regular position is 
not uniform; 
displacement vectors display  local structures resembling vortexes of incompressible fluid.  
Here again, it is concluded that 
the local displacement of closely packed organs 
gives rise to the systematic variations of 
$r_n$ and divergence angle $d_n$. 

%
%
%




%

%

\section{Discussion}


Discontent not only with untested abstractions in the  mathematical literature 
but also with vague expressions 
like ``the largest available space'' and ``about 137.5$^\circ$''
in the botanical literature 
underlies this work. 
It is often assumed implicitly 
that a spiral pattern has a definite center.  
This is not necessarily true, as remarked in Sec.~\ref{sec:floatingcenter}. 
As shown in Sec.~\ref{sec:floatingcenterP},  
each cell of a parastichy lattice may have its own center. 
The center of parastichy is not fixed in space, but 
it may float around in the pattern as primordia arise one after another.  
The floating center can be the main source of apparent variations of
divergence angle.  
%
In the literature, 
a logical leap is often made 
from a measured value of about 137.5 to the irrational number $360/\tau^2$ 
without subjecting data to statistical analysis. 
The lack of quantification is partly because 
the center against which to measure the angles is uncertain. 
Probably for this reason, 
little attention has been directed to 
the numerical accuracy of divergence angle. 
%
%
%
%
%
%
%
%
%
%
%
%
The standard deviation of divergence angle 
conveys no less important information than the mean value.  
According to an observation, 
the standard deviation 
decreases systematically  as the plant grows day by day
while the mean is kept constant (\cite{williams74}). 
This behavior is not confirmed by snapshot patterns investigated in this paper. 
The present work finds no significant drift of  the mean divergence angle, 
which puts possible mechanisms of phyllotaxis under constraint. 

There are two cases where it is invalid to assume for 
a phyllotactic pattern to have a fixed center. 
In one case, the position of center is displaced as a leaf primordium 
is initiated. 
In the other case, leafy organs are displaced 
by secondary effects after they are initiated at their regular position. 
The effects of the former and the latter are taken into account 
by means of the floating centers
in Sec.~\ref{sec:floatingcenter} and 
Sec.~\ref{sec:floatingcenterP}, respectively. 
It is noted that 
the former concept conforms with observation by \cite{gb87}, 
who have made a general remark that the center of apical area moves as
a new leaf is (or leaves are) initiated. 
Their observation is verified concretely by the methods and results of this paper. 
\cite{gb87} note that a trajectory of the moving center
may be used as a diagnostic for phyllotactic patterns,  which is to be contrasted
with classification by means of geometrical unit of ideal patterns (\cite{mz89}).


At a glance of an impressive phyllotactic pattern, 
our attention is apt to be directed to Fibonacci numbers, because our eyes tend to follow conspicuous parastichies. 
However, parastichies do not serve much for quantitative purposes. 
Polar coordinates $(r_n, \theta_n)$ of pattern units 
cannot be deduced unequivocally from parastichy numbers alone, although 
parastichy numbers are determined uniquely from coordinates $(r_n, \theta_n)$. 
The polar coordinates are the most proper tool to specify a
two-dimensional pattern quantitatively, 
only with which it can be investigated whether and how 
the divergence angle $\theta_n-\theta_{n-1}$ 
and the radial coordinate $r_n$ depend on the leaf index $n$. 
If a pattern is deformed to lose a well-defined center, 
the meaning of divergence angle blurs apparently. 
%
Still, parastichies may remain little affected owing to their insensitivity to
quantitative details. 
Even an oval-shaped capitulum preserves parastichy numbers (\cite{sp94}). 
The method of \ref{how2number} for indexing primordia 
remains valid irrespective of whether the pattern has a center or not, 
because it is based solely on parastichies. 
Although a real system generally do not possess a well-defined center,
the method of the floating centers 
enables us to infer coordinates $(r_n, \theta_n)$ of the original
pattern,  
as shown in the last section. 
%

\begin{figure}[t]
\centering
\subfigure[]{
\includegraphics[width= .45\textwidth]{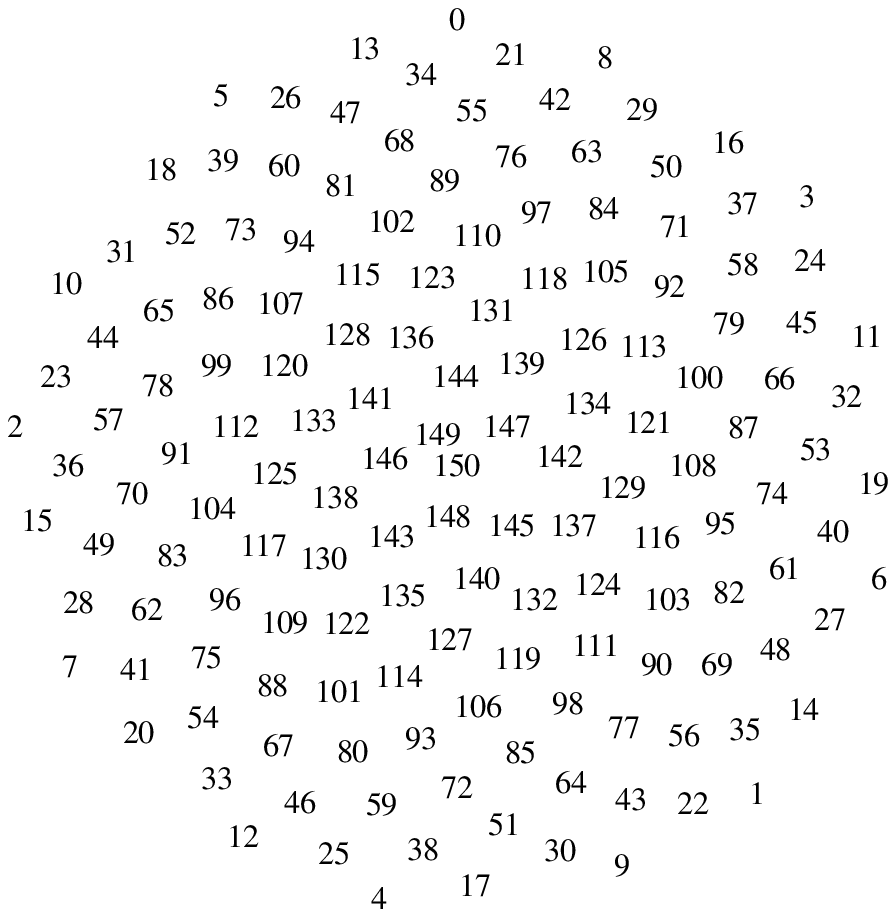}
\label{realphyllo}
}
\subfigure[]{
\includegraphics[width= .45\textwidth]{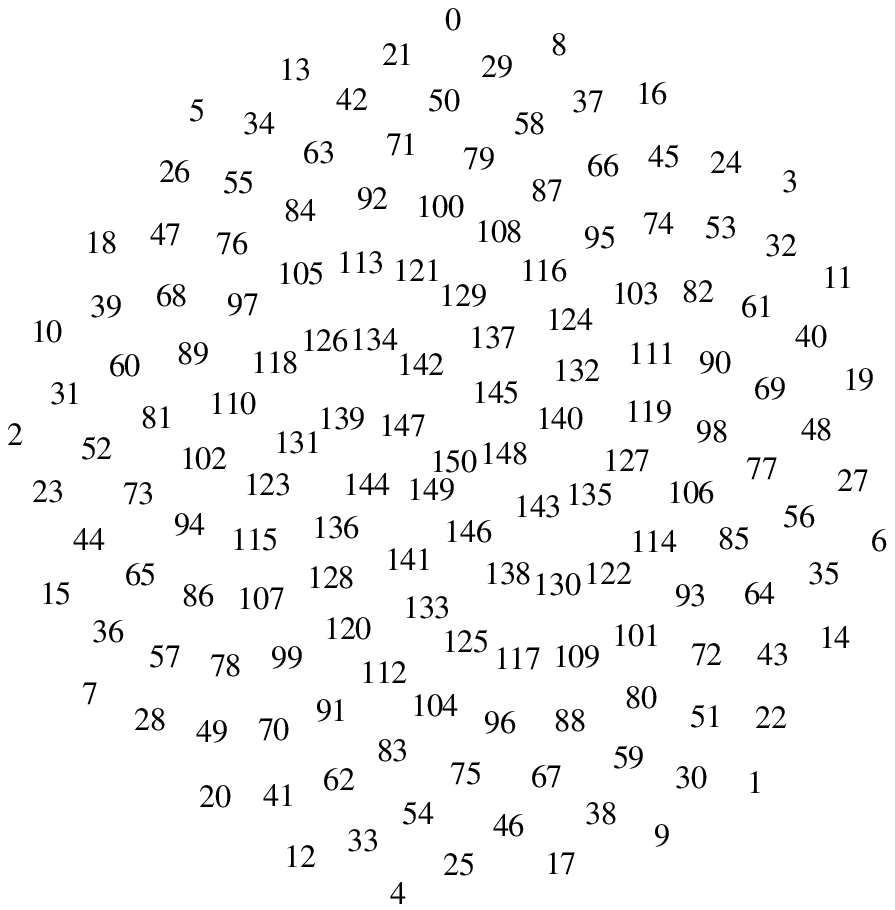}
}
\caption{
Real and fake phyllotaxis to illustrate accurate control of divergence angle~$d$. 
Densely packed florets arise spirally in numerical order {from the rim}
 toward the center.
Inner florets may be omitted without affecting arrangement near the rim,
 that is at issue. 
(a) Normal phyllotaxis of a $(21,34)$ pattern for $d=137.5^\circ$,
 quite common in nature. 
(b) Seemingly plausible but actually improbable phyllotaxis of 
a $(21,29)$ pattern for $d=136.8^\circ$. 
Plants distinguish the difference of less than 1$^\circ$ without fail, 
whereas the human eye would find it difficult to tell (a) from (b).  
The difference comes to the fore when the 21st floret arises on either
 side of the 0th floret;  plants know the correct position (angle) in advance. 
Central portions after, say, the 120th floret 
are overlapped nearly perfectly by a rotation of
 about 90$^\circ$, so that they are indistinguishable by means of lower
 parastichy numbers there.
\label{sunflower}
}
\end{figure}
The results of this work underline the universal characteristic 
of phyllotaxis:  
divergence angle during steady growth 
is stably held at $137.5^\circ$, or rarely $99.5^\circ$
and other special values. 
This is not just a general tendency. 
As a pivotal empirical rule deduced from experimental facts
that lays the foundation of mathematical analysis
and thus underlies this intriguing phenomenon of phyllotaxis, 
the author believes that it is appropriate to 
attribute the status of a natural law to the universality of 
the accurate control of divergence angle
(cf. \ref{sec:logarithmicspiral}). 
%
To convince that this is not exaggeration, 
let us take two examples. 
The first is given by high phyllotaxis of florets on a capitulum. 
According to (\ref{p'q'alpha0pq}), a $(21,34)$ pattern is obtained if and only if 
divergence angle $d$ is constrained within a narrow range of 
$\frac{8}{21}<\frac{d}{360^\circ}<\frac{13}{34}$, 
or $137.14^\circ < d< 137.65^\circ$. 
Imagine what happens if divergence angle had 
been 
out of 
the range of width of 0.5$^\circ$. 
As a mathematical consequence, it is shown that 
$(21,29)$ or $(13,34)$ should obtain instead of the normal phyllotaxis $(21,34)$
(Fig.~\ref{sunflower}). 
Neither of these anomalous patterns is observed actually. 
Thus, 
not only the constancy of divergence angle 
but also the accurate control of it to the special value of $137.5^\circ$ 
should be taken seriously as an empirical fact. 
The point is not just that a normal $(21,34)$ phyllotaxis {\it is}
observed, but rather that anomalous patterns of 
$(21,29)$ and $(13,34)$ phyllotaxis do {\it not} coexist, or mixed, with
the normal pattern. 
%
%
%
Thus, plants must know precisely
the correct value of divergence angle  beforehand. 
The second example is given by low phyllotaxis of young leaves on the apex. 
As shown in Sec.~\ref{sec:suaedavera}, 
divergence angle is 
independent of sizes of leaves. 
This is an important observation, 
because it apparently conflicts with the assumption of dynamical models that 
divergence angle is variable depending on the size of leaves (\cite{mbvb12}). 
So to speak, the whole is more than the sum of its parts. 
It seems rather appropriate to regard divergence angle as a given constant parameter  (\cite{src12}).
Along with a review on various approaches to phyllotaxis, a model conforming to
these observations has been given previously (\cite{okabe12}). 
%
%
%
%
%
%
%
%


A remark is in order. 
In the ontogeny, divergence angle generally begins with 180$^\circ$ or 90$^\circ$ of a distichous or decussate pattern, 
and thereafter shows some transient fluctuations before attaining toward a universal constant value. 
Variations in the transient regime are substantially larger (typically 
more than about $10^\circ$) 
than the small variations in the steady-growth regime investigated throughout this paper. 
Therefore, the two variations are distinguished quantitatively. 
The former may be understood as a natural result of 
the limited numbers of apical meristem cells pushing and shoving with each other. 
The latter variations, however, are too small to be explained as such; 
actually, as remarked above,  it appears rather that
divergence angle is regulated steadily at a predetermined value, despite any adventitious factors. 
Generally speaking, living systems are so complex that large variations are by far easier to understand. 
For a quantitative
 understanding, 
the riddle of the small variations cannot be overemphasized. 
It should be investigated for its own sake, separately from the large
transitory fluctuations during ontogeny. 


%
%




%





It is also a finding of this paper that apparent deviations from the universal mean 
are not due to random errors as commonly expected, but they are actually
 systematic variations. 
On the one hand, 
 they are likely to be due to directional correlation with respect to the sun (\cite{kumazawa71}).
%
For lack of space, it is only noted here that
the azimuthal correlation, as plotted in Fig.~\ref{rutishauser:subfig3},
is a common phenomenon confirmed quite generally.  
On the other hand, 
systematic variations are caused also by local displacement when organs are closely packed.  
Concerning the latter,
\cite{rrb91} have attached importance to 
persistent fluctuations of individual divergence angles (open circles in Fig.~\ref{rrb4:subfig2})
as a real phenomenon. 
%
%
%
As indicated in Figs.~\ref{rrb4:subfig3} and \ref{rrb4:subfig8}, 
the fluctuations systematically correlated with the azimuth, the
absolute direction, 
should be regarded as a subsidiary phenomenon.  
Thus, one should be cautious not to take 
apparent fluctuations 
literally as evidence for or against a theoretical model of phyllotaxis. 
This is an important point because 
almost all experimental and theoretical 
divergence angles reported so far 
have not taken into account the possibility that the center for the
angles is not fixed in space. 
Therefore, virtually all the reported results  have to be reexamined carefully. 
If variations turn out to be due to misidentification of the center, 
they have no real significance.  
If they reflect real effects, they act to disturb
the regularity.
The disturbance, if any, conveys important information,   
which may be analyzed with a model from an appropriate theoretical perspective. 

As illustrated by Figs.~\ref{rrb4vector} and \ref{hotton:subfig9},  
variations of divergence angle are interpreted naturally as necessary
consequences of 
local distortion due to contact pressure between closely packed organs.
The contact pressure may affect phyllotaxis 
not only quantitatively but even qualitatively. 
Primordia under high pressure from the surrounding primordia may fail to grow normally, 
as illustrated by innermost florets on a capitulum in Fig.~\ref{rrb4vector}.  
In general, organs in a crowded portion 
are most likely to be aborted, or obliged not to arise normally. 
Ill effects due to an aborted primordium may propagate along a parastichy of vascular connection,
 along which nutrients are transported. 
The other possibility is that new primordia may happen to arise to fill in gaps 
opened between existing primordia. 
The author suspects that aborted or inserted primordia 
might appear as a crystallographic defect in an otherwise regular
parastichy lattice (\cite{zm94}). 
\cite{sp94} has made an interesting observation that 
the parastichy number decreases quite more often than it increases.

Individual divergence angles measured against a nominal center
may appear to vary very wildly. 
Nonetheless, it has been 
empirically known  that 
the limit divergence angle of 137.5$^\circ$ is immediately obtained 
 by evaluating the average of several successive angles covering a few full turns.  
%
%
%
%
%
%
This empirical fact 
is explained consistently by the observation of the present work 
that the apparent variations are not random but systematically 
correlated with the azimuth. 
If the variations are of random origin, 
the standard deviation should depend on the number of sample leaves. 
The present study has found no such dependence. 
%
In contrast, the systematic variations correlated with the azimuth are averaged away within a few full turns of the azimuth.   
For the samples analyzed in this paper, 
the systematic variations are of the order of 1~degree. 
If this secondary effects of the directional correlation 
are compensated for, 
the accuracy and stability of angular control by nature should stand out
all the more strikingly; 
it should turn out to be of the order of 0.1~degree as a general rule. 
This is an unexpected result. 
As remarked above, 
the numerical accuracy of this quantitative phenomena of living organs 
should arrest more
 attention than eye-catching Fibonacci integers do. 

As shown by the results, 
the angular equation (\ref{thetan=nd}) with constant divergence
angle $d$ remains valid quite generally. 
In contrast, the radial equation (\ref{rn=Asqrtn}) is not valid
quantitatively, 
whereas (\ref{rn=an}) holds true at the shoot apex in a good approximation. 
Unlike the divergence angle $d$, however, 
the plastochron ratio $a$ may be changed artificially (\cite{me77})
(Sec.~\ref{egaonxanthium}). 
As a matter of fact, 
it has been reported that the plastochron ratio $a$ changes naturally during shoot development. 
In particular, the ratio decreases appreciably during transition to
flowering without changing the phyllotactic sequence determined by the
divergence angle $d$. 
These observations suggest that 
the growth in the radial direction $r_n$ is controlled 
independently of the angle $\theta_n$ 
(cf. (\ref{ddnlogrn=loga}) and (\ref{ddnpirn2})). 
In other words, 
the regularity in $r_n$ may not be dealt with 
on the same basis as the regularity in $\theta_n$. 
Indeed, the present work finds no correlation between the angle
$\theta_n$ and the radius $r_n$. 
This point has bearing on how to determine 
the numerical order of leaves, namely the leaf index system $n$. 
Two index systems based on the numerical orders of $r_n$ and $\theta_n$ need
not be identical. 
Owing to the regularity in $\theta_n$, 
the index $n$ is orderly set according to the angle $\theta_n$.
%
In contrast, the radius $r_n$ 
in practice is not a monotonic function of $n$. 
Fig.~\ref{rutishauser:subfig4} indicates that $r_n$ decreases non-monotonically, 
or the order in $n$ (horizontal axis) is not preserved for $r_n$ (vertical axis). 
Actually, this is normally the case when the plastochron ratio $a$ is very close to 1 (high phyllotaxis), for 
irregular fluctuations in $r_n$ outweigh regular changes by the plastochron ratio $a$. 
Even the height order
 of successive leaves on a stem may be interchanged (\cite{fujita42}).   


Excepting such minor irregularities, which are more or less expected for
living organisms,  
observations generally support validity 
of {mathematical} description in terms of 
the polar coordinate system. 
This is not at all trivial. 
Just as an elliptic orbit of a planet pins down a special point in space, the sun at the origin,   
the general regularities manifested plainly by means of
the polar coordinate system signify the existence of a special singular point, 
the origin of the coordinate system.   
Unlike the solar system, there is no obvious sign at the origin
of a phyllotactic pattern. 
Indeed this was a motivation of the present work; even without assuming the center, 
a point of the sort has been indicated definitely (see Fig.~\ref{me2:subfig1}). 
The origin or the center
should be identified with 
the growing point biologically, though it is meant here 
in a narrower sense than in general botanical use. 
The author speculates that 
biochemical properties of the singular point, the growing center of a spiral pattern, 
  might hold a key to understanding 
not only mechanisms of radial growth but also 
regulation mechanisms of divergence angle.  

%
%
%

\appendix
\section{Logarithmic spiral and parastichy concept revisited}
\label{sec:logarithmicspiral}

Logarithmic spirals in phyllotaxis has been investigated intensively 
by \cite{church04}, \cite{vaniterson07}, \cite{richards51}, \cite{erickson83} and \cite{jean94}, among others. 
The main purpose of this section 
is to derive the formulas used in the main text,  
(\ref{loga=2pisqrt}), (\ref{Jloga=2pisqrt}), (\ref{loga2pisqrt5tau2n+1}) and
(\ref{Jloga2pisqrt5tau2n+1}) for the plastochron ratio $a$
when two opposite parastichies are orthogonal, 
as they have not been presented before in these forms.
The following derivation has a merit
of transparency by which a physical assumption and a mathematical approximation 
are distinguished
by excluding unnecessary assumptions and complications. 
As remarked below, this distinction is essential to a clear understanding of 
empirical rules of phyllotaxis and Richards' phyllotaxis index.  
%
Before that, the prevalent concept of {\it parastichy} must be made clear and definite.
It is also an important aim of this section to point out that 
conventional expositions are incorrect or insufficient. 
%
%
%



In the polar coordinate system $(r,\theta)$, 
a logarithmic spiral through a point $(r,\theta)=(1,0)$ is given 
by 
\begin{equation}
r= e^{b\theta},    
\label{rr0expbtheta}
\end{equation}
where $b$ is a constant. 
The logarithmic spiral is characterized by the property that
the angle $\varphi$ 
between the radial vector from the origin and the tangent vector 
is held constant at every point $(r, \theta)$ on the spiral.   
\begin{equation}
 r\frac{d\theta}{dr}=\tan \varphi. 
\end{equation}
Hence it is also called the equiangular spiral. 
The angle $\varphi$ is related to the coefficient $b$ in
(\ref{rr0expbtheta}) by 
\begin{equation}
 b=\cot\varphi.   
\label{b=cotvarphi}
\end{equation}


The fundamental spiral of phyllotaxis given by (\ref{thetan=nd}) and (\ref{rn=an})
forms a logarithmic spiral with 
\begin{equation}
 b= \frac{\log a}{d}=\frac{\log a}{2\pi \alpha_0}.  
\label{b=logad=}
\end{equation}
In the second equation, 
a reduced divergence angle $\alpha_0$ 
is introduced by 
\begin{equation}
 d=2\pi \alpha_0. 
\label{d=2pialpha0}
\end{equation}
Owing to the periodicity in angle, 
one may assume 
either $0\le \alpha_0 < 1$ or $-\frac{1}{2}< \alpha_0 \le \frac{1}{2}$ 
without loss of generality. 
In the latter convention, the direction of the fundamental spiral is
determined by the sign of divergence angle. 
In what follows, 
it is assumed $0\le \alpha_0 \le \frac{1}{2}$, 
 i.e., the fundamental spiral runs outward counterclockwise.  
It is straightforward to adapt the following results 
to spirals for 
$-\frac{1}{2}< \alpha_0 \le 0$. 

%
%
%

%
%

The concept of {\it parastichy} is widely used in the literature. 
As pointed out explicitly below, 
prevalent expositions are not satisfactory.
According to a general consensus, 
a spiral connecting 
leaves with the indices differing by an integer $q$ is referred to as a
$q$-parastichy. 
However, there are infinitely many such spirals. 
Therefore, the term {\it  $q$-parastichy} is not appropriate in the first place. 
%

As a $q$-parastichy, 
consider a logarithmic spiral $r= e^{b\theta}$
running through the 0-th and $q$-th leaves, 
namely
$(r,\theta)=(1,0)$ and $(a^q, 2\pi q \alpha_0)$. 
The former is met from the outset, whereas 
the latter determines the coefficient $b$. 
Owing to the periodicity in angle, 
the polar coordinates of the $q$-th leaf
$(r_q,\theta_q)=(a^q, 2\pi q \alpha_0)$ is equivalently represented as
 $(a^q, 2\pi q \alpha_0- 2 \pi p)$,  
where $p$ is an arbitrary integer. 
Substituting the latter into the spiral equation $r=e^{b\theta}$, 
\begin{equation}
 b= \frac{\log a^q}{2\pi  \left(q \alpha_0- {p}\right)} 
= \frac{\log a}{2\pi \left(\alpha_0- \frac{p}{q}\right)}. 
\label{b}
\end{equation}
As the spiral thus determined runs through all leaves with the index $n$
divisible by the integer $q$, 
it is 
a $q$-parastichy. 
The $q$-parastichy depends on the integer $p$. 
There are as many $q$-parastichies as the number of integers. 
The fundamental spiral with (\ref{b=logad=})
is a special case of (\ref{b}) for $(p,q)=(0,1)$. 
%
%
%
%
%
In the literature, the most conspicuous, shortest one is 
usually regarded as the $q$-parastichy. 
Therefore, 
the integer $p$ is identified with the integer nearest to
$q\alpha_0$.
As a matter of fact, this need not 
be the case. 

In practice, it is necessary and sufficient to let $p$ be 
an integer approximating a number $q\alpha_0$. 
The denominator of the middle expression of (\ref{b}) represents 
the net angle between two successive leaves on the $q$-parastichy. 
To keep the angle within a reduced range of width $2\pi$, 
a multiple of a full turn,  $2\pi {p}$, is subtracted from 
the nominal angle of $2\pi q \alpha_0$. 
%
Hence the integer $p$ is 
the winding number 
of the fundamental spiral executed 
between consecutive leaves on  the $q$-parastichy. 
%
If the number $q\alpha_0$ is not an integer, 
 $p$ is most properly chosen to be 
either the largest previous or the smallest following integer
to $q\alpha_0$. 
In the former case, $q\alpha_0-p>0$, 
the $q$ parastichy spirals 
in the same direction as the fundamental spiral. 
In the latter case, $q\alpha_0-p<0$, 
the direction of the $q$ parastichy is opposite to the fundamental spiral. 
If $q\alpha_0$ is an integer, then $p=q\alpha$,  
and $\varphi=0$ by (\ref{b=cotvarphi}). 
In this special case, the parastichy is not a spiral curve but 
a straight half-line radiating from the center. 
Therefore it is especially called the orthostichy.
It goes without saying that 
orthostichy in the strict sense does not exist in real life, as 
it does not make sense to ask whether $q\alpha_0-p$ is zero or not; 
it is equivalent to asking whether $\alpha_0$ is an irrational or a rational
number.  
In any case, 
{\it the parastichy depends not only on the parastichy number $q$
but also
on the winding number $p$}, or more precisely 
on $\alpha_0-\frac{p}{q}$ whose sign and magnitude 
uniquely determine the sense and slope angle of the parastichy. 

\begin{figure}[t]
\begin{center}
\includegraphics[width=0.5\textwidth]{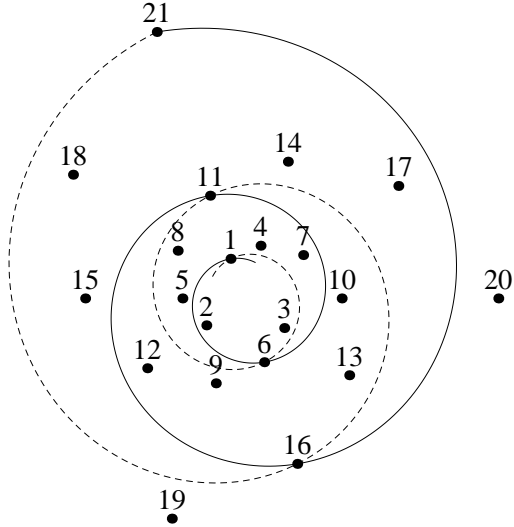}
\caption{
A phyllotactic pattern for $\alpha_0=\frac{3}{10}$ and $a=1.1$. 
Two distinct 5-parastichies connect the same set of points, 1, 6, 11,
 16, 21; 
the solid curve is a 5-parastichy for $(p,q)=(1,5)$, 
and the dashed curve is 
another 5-parastichy
for $(p,q)=(2,5)$. 
The latter spiral winds in the direction opposite to the fundamental spiral. 
\label{alpha310}
}
\end{center}
\end{figure}

As the parastichy, or (\ref{b}),  depends only on the fraction $\frac{p}{q}$, 
it is sufficient to consider the {\it irreducible} pair of $p$ and $q$. 
That is to say, if $p=0$, then $q=1$; 
otherwise, $p$ and $q$ are restricted to 
coprime 
integers, i.e., $p$ and $q$ are not evenly divided by any integer
greater than 1. 
%
%
%
%
%
%
By the assumption $0\le \alpha_0 \le \frac{1}{2}$, 
it is sufficient to consider 
irreducible fractions between 0 and $\frac{1}{2}$. 
For the denominator $q$ up to 8, they are
\begin{equation}
\frac{p}{q} = 
\frac{0}{1}, 
\frac{1}{2}, 
\frac{1}{3}, 
\frac{1}{4}, 
\frac{1}{5}, \frac{2}{5}, 
\frac{1}{6}, \frac{1}{7}, \frac{2}{7}, \frac{3}{7}, 
\frac{1}{8}, \frac{3}{8}. 
\label{pq8}
\end{equation}
Hence, parastichies for $q=5,7,8$ are not unique. 
By way of illustration, 
five parastichies for $\frac{p}{q} = \frac{1}{5}$ and $\frac{2}{5}$
are shown in Fig.~\ref{alpha310} for $\alpha_0=\frac{3}{10}$.  
It is not common to notice 5-parastichies in the pattern of
Fig.~\ref{alpha310}. 
This is because the pattern has more conspicuous parastichies.  
A numerical measure of conspicuousness is given by the
absolute value of $\alpha_0-\frac{p}{q}$, 
which is a winding angle per leaf of the parastichy. 
In a word, the less winding the parastichy is, the more conspicuous it appears to the eye. 
In other words, {\it the better 
the fraction $\frac{p}{q}$ approximates the divergence angle $\alpha_0$, 
the more conspicuous the parastichy is. }
To find conspicuous parastichies for a given value of $\alpha_0$, 
a sequence of irreducible fractions
arranged in the numerical order, called a Farey sequence, is of much help. 
By arranging (\ref{pq8}),
we obtain the Farey sequence of order $q=8$; 
\begin{equation}
\frac{0}{1}, 
\frac{1}{8}, \frac{1}{7}, \frac{1}{6}, \frac{1}{5}, \frac{1}{4},
 \frac{2}{7}, \frac{1}{3}, \frac{3}{8}, \frac{2}{5}, \frac{3}{7}, 
\frac{1}{2}. 
\label{farey8}
\end{equation}
As $\alpha_0=\frac{3}{10}$ lies between 
$\frac{2}{7}$ and $\frac{1}{3}$ in this sequence, 
it is properly understood that 3 and 7 parastichies 
make a conspicuous pair for the pattern of Fig.~\ref{alpha310}.


In spite of the above remark, 
the conventional term of the $q$-parastichy is used throughout
this paper 
in order not to complicate matters by introducing a new term 
like a $^pq$-parastichy. 
In practice, we get along without encountering 
the ambiguity because the number $q\alpha_0$ always happens to be 
nearly an integer, namely a Fibonacci number,
 by the empirical fact that 
divergence angle $\alpha_0$ is given by a special irrational number
(cf. (\ref{alpha0=tau-2})). 
Therefore, the integer is always identified with $p$ unwittingly. 
Another reason is that 
different sets of parastichies with a common parastichy number never
become conspicuous at the same time, 
because adjacent terms in a Farey sequence have different denominators (\cite{hw79}).
This is a consequence of (\ref{pq'-p'q=1}) and concretely checked by
(\ref{farey8}).

The second shortest $q$-parastichy may become relevant. 
%
For instance, consider a case for $\alpha_0=\frac{1}{12}$ and $q=3$ (\cite{swinton12}).  
The 3-parastichy with $p=0$ is actually not a 3-parastichy because $(p,q)=(0,3)$ is
reducible to $(0,1)$, the fundamental spiral. 
Therefore, a parastichy for $(p,q)=(1,3)$ should rather be referred to as the 3-parastichy. 
%
%
%
%
%
%
%
%
%
This example refutes the prevalent statement
that $p$ 
is {the integer nearest to} $q\alpha_0$ (\cite{jean94,swinton12}).


Along with the $q$ parastichy with $b$ in (\ref{b}), 
let us consider a $q'$-parastichy for another arbitrary integer $q'$. 
In a similar manner as above,  
a $q'$-parastichy is given by a logarithmic spiral with 
\begin{equation}
 b'= \frac{\log a}{2\pi \left(\alpha_0- \frac{p'}{q'}\right)}   
\end{equation}
for $b$ in (\ref{rr0expbtheta}), 
where $p'$ is an integer approximating $q'\alpha_0$. 
According to (\ref{b=cotvarphi}),  
the coefficient $b$ and $b'$ for the $q$ and $q'$ parastichy 
are related with the slope angle $\varphi$ and $\varphi'$ 
by $b=\cot\varphi$ and $b'=\cot\varphi'$, respectively. 
When the $q$-parastichy and the $q'$-parastichy are mutually orthogonal, 
%
%
%
%
the identity $\varphi+\varphi'=\frac{\pi}{2}$ holds true. 
Accordingly,  
$bb'=\cot \varphi\cot \varphi'=-1$, i.e., 
\begin{equation}
\frac{\log a}{2\pi \left(\alpha_0- \frac{p}{q}\right)}  
\frac{\log a}{2\pi \left(\alpha_0- \frac{p'}{q'}\right)} =-1, 
\end{equation}
or
\begin{equation}
\log a=2\pi
 \sqrt{\left(\frac{p}{q}-\alpha_0\right)\left(\alpha_0-\frac{p'}{q'}\right)}, 
\label{loga=2pisqrt}
\end{equation}
%
which is real and positive. 
The inequality $bb'<0$, which holds when the two parastichies are opposite, 
signifies that 
$\alpha_0$ should lie in between the 
two fractions $\frac{p}{q}$ and $\frac{p'}{q'}$. 
Let the former fraction be numerically larger than the latter,  without loss of generality. 
Then,  
\begin{equation}
 \frac{p'}{q'}< \alpha_0 < \frac{p}{q}. 
\label{p'q'alpha0pq}
\end{equation}

In the case of $\alpha_0<0$, 
both $p$ and $p'$ become not positive, so that the
 inequalities in (\ref{p'q'alpha0pq}) should be reversed, 
whereas (\ref{loga=2pisqrt}) remains valid formally as it is.  

\begin{figure}
\begin{center}
\includegraphics[width=0.5\textwidth]{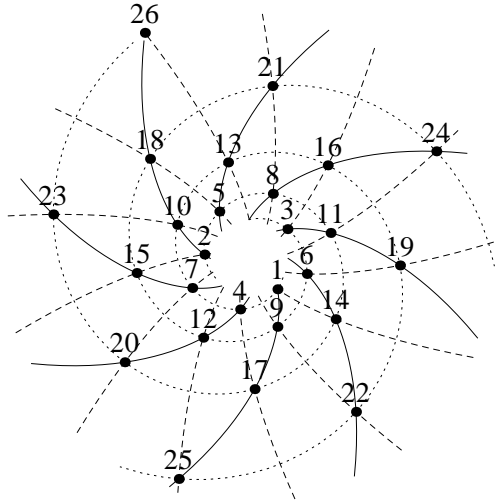}
\caption{
A phyllotactic pattern with divergence angle 
$\alpha_0=\tau^{-2}$ and the plastochron ratio $a$ in 
(\ref{loga=2pisqrt}) for $\frac{p}{q}=\frac{5}{13}$ and
 $\frac{p'}{q'}=\frac{1}{3}$. 
This pattern has an orthogonal parastichy pair of $(q,q')=(3,13)$, 
i.e., 3-parastichies (dotted) and 13-parastichies (dashed) cross orthogonally. 
Nonetheless, 
it is usually referred to as a $(8,13)$  phyllotaxis, 
because 8-parastichies (solid) with $\frac{p''}{q''}=\frac{3}{8}$ are as conspicuous as
 13-parastichies.   
As $pq''-p''q=1$ for $(q'',q)=(8,13)$ whereas $pq'-p'q=2$ for
 $(q',q)=(3,13)$, 
the former has a reason to be regarded as the fundamental pair. 
Parastichies are imaginary constructs, 
though some 
 may have real observable effects.
\label{SAM}
}
\end{center}
\end{figure}
A pair of parastichies 
is commonly referred to by the denominator pair $(q,q')$.  
The parastichy pair $(q,q')$ 
is usually not chosen so that the parastichies are nearly orthogonal; 
it is chosen to be {\it irreducible}, i.e., such that 
the two limiting fractions in (\ref{p'q'alpha0pq}) constitute successive
terms of a Farey sequence.  See Fig.~\ref{SAM}.  
By a theorem of number theory (\cite{hw79}), 
it is equivalent to stating that 
integer pairs for a pair of parastichies $(p,q)$ and $(p',q')$ 
are chosen to satisfy 
\begin{equation}
 pq'-p'q=1. 
\label{pq'-p'q=1}
\end{equation}
The identity is confirmed for adjacent fractions in (\ref{farey8}). 
The {\it irreducible} pair of parastichies 
should not be confused with 
the {\it irreducible} numbers $(p,q)$ for a $q$-parastichy discussed above. 
The former may be rephrased as {\it conspicuous}, 
although the term carrying ordinary connotations might better be avoided. 
In Fig.~\ref{SAM}, the parastichy pair $(3,13)$ is not conspicuous (irreducible),
whereas not only $(8,13)$ but $(3,5)$, $(5,8)$, $(13,21)$
and so on are conspicuous (irreducible).  
Thus, 
(\ref{pq'-p'q=1}) should be considered as 
a convenient prescription for good choices of the parastichy pair.  
%
Physically, 
the left-hand side of (\ref{pq'-p'q=1}) 
represents the number of leaf points in a unit cell of the lattice spanned by $q$ and $q'$ parastichies (Fig.~\ref{SAM}).  
(\ref{pq'-p'q=1}) means not only that $q$ and $q'$ are 
coprime integers but that $p$ and $p'$ are determined independently of $\alpha_0$. 
As a result, the parastichy numbers do not depend on the divergence angle, 
insofar as the condition (\ref{p'q'alpha0pq}) is met. 
Note that the phyllotactic pattern is completely characterized by the 
two parameters, the plastochron ratio $a$ 
and the divergence angle $\alpha_0$, 
which are not determined uniquely by parastichy numbers alone. 
The relation between the divergence angle $\alpha_0$ and the limiting fractions $\frac{p}{q}$ and $\frac{p'}{q'}$ 
to be used for (\ref{loga=2pisqrt}) 
is immediately read from a diagram presented in \cite{okabe11,okabe12}.

%
%
%
%


An opposite parastichy pair
of a multijugate system has  the greatest common divisor $J >1$, 
so that the pair is written $J(q, q')$, 
where $q$ and $q'$ are 
coprime integers. 
Instead of (\ref{loga=2pisqrt}),
 one obtains
\begin{equation}
\log a=\frac{2\pi}{J}
 \sqrt{\left(\frac{p}{q}-J\alpha_0\right)\left(J\alpha_0-\frac{p'}{q'}\right)}, 
\label{Jloga=2pisqrt}
\end{equation}
where $p$ and $p'$ are integers near 
$J\alpha_0 q$ and $J\alpha_0 q'$. 
The divergence angle $\alpha_0$ of the $J$-jugate system 
satisfies 
\begin{equation}
 \frac{p'}{q'}< J\alpha_0 < \frac{p}{q}. 
\end{equation}
If  $J(q, q')$ is 
a conspicuous pair, 
\begin{equation}
 Jq' p -Jqp'  =J, 
\label{pq'-p'q=J}
\end{equation}
or (\ref{pq'-p'q=1}) holds true.  
Thus, 
the divergence angle $\alpha_0$ and the plastochron ratio $a$ of a
multijugate system may be better expressed in the forms of $J\alpha_0$
and $a^J$ to compare them with a single spiral system 
(Sec.~\ref{sec:trijugate}). 


%
%
%
%
%
%
In the above, 
no approximation has been made 
besides the basic equations 
(\ref{thetan=nd}) and (\ref{rn=an}). 
Above all, the derivation is not based on 
any ideal assumption on leaf shape, e.g., 
{\it contiguous circles} abundant in the phyllotaxis literature. 
This is practically of crucial importance because any argument based on `perfectly circular leaves'
must meet with severe criticism from experimentalists. 
As a matter of fact, 
parastichy is a property of a point lattice, 
so that it is independent of shape and size of pattern units. 
If (\ref{pq'-p'q=1}) is to be regarded as the condition for 
a {\it visible} parastichy pair, 
this is what \cite{jean94} calls the
fundamental theorem of phyllotaxis. 
But then, 
one must be careful about 
what is {\it visible} (\cite{swinton12}). 
See a remark below (\ref{pq'-p'q=1}) on 
the term {\it irreducible}.

The above formulation is general. 
%
According to number theory, 
linear Diophantine equations, (\ref{pq'-p'q=1}) and (\ref{pq'-p'q=J}), 
are solved for any parastichy pair of integers.
%
If the parastichies are orthogonal, the plastochron ratio $a$ is given by 
(\ref{loga=2pisqrt}) or (\ref{Jloga=2pisqrt}). 
Nevertheless, 
nature adopts only selected numbers as the parastichy pair. 
Parastichy numbers of a normal phyllotaxis comprise 
the Fibonacci
sequence,  
\begin{equation}
F_i=1, 1, 2, 3, 5, 8, \cdots
\end{equation}
for 
\[
i=1, 2, 3, 4, 5, 6, \cdots, 
\]
respectively. 
The Fibonacci numbers satisfy 
the mathematical identity 
\begin{equation}
F_{ i-1} F_{ i}-F_{i-2} F_{i+1}  = (-1)^i.  
\end{equation}
Therefore, a solution of (\ref{pq'-p'q=1}) is given by
\begin{equation}
(q,q' ,p, p')=(F_{i+1}, F_{i}, F_{i-1}, F_{i-2})
\end{equation} 
if $i$ is an even integer, 
or by 
\begin{equation}
(q,q',p,p')=(F_i, F_{i+1},F_{i-2}, F_{i-1})
\end{equation}
if $i$ is odd. 
%
%
%
%
%
In either case, 
(\ref{loga=2pisqrt}) gives 
\begin{equation}
\log a=2\pi
 \sqrt{\left(\frac{F_{i-2}}{F_i}-\alpha_0\right)\left(\alpha_0-\frac{F_{i-1}}{F_{i+1}}\right)} 
\label{loga=2pifn-2fn-alpha0}
\end{equation}
for the orthogonal Fibonacci parastichy system of $(F_i, F_{i+1})$. 
(\ref{loga=2pifn-2fn-alpha0}) is valid insofar as $\alpha_0$ lies between 
$\frac{F_{i-2}}{F_i}$ and $\frac{F_{i-1}}{F_{i+1}}$. 
Nonetheless, it is usually taken for granted that 
divergence angle
$\alpha_0$ is identified with the mathematical limit of $\frac{F_{i-2}}{F_i}$ and
$\frac{F_{i-1}}{F_{i+1}}$ as $i\rightarrow \infty$, that is, 
\begin{equation}
 \alpha_0=\frac{1}{2+\tau^{-1}}= \tau^{-2}
\label{alpha0=tau-2}
\end{equation}
 (\cite{richards51,jean83}), 
where $\tau$ is the golden ratio given in (\ref{tau}). 
The divergence angle of $d\simeq 137.5^\circ$ for (\ref{alpha0=tau-2}) 
is called the limit divergence angle of the main (Fibonacci) sequence.  
To show that (\ref{alpha0=tau-2}) is the limit of $\frac{F_{i-2}}{F_i}$, 
the fraction is expanded with respect to $\tau^{-1}$ after 
the exact formula 
\begin{equation}
{F_i} =\frac{\tau^i-(-\tau)^{-i}}{\sqrt{5}}
\end{equation}
is substituted for the denominator and numerator; 
\begin{equation}
 \frac{F_{i-2}}{F_i} =\frac{\tau^{i-2}-(-\tau)^{-i+2}}{\tau^i-(-\tau)^{-i}} 
\simeq \tau^{-2} - (-1)^{i} \sqrt{5}\tau^{-2i}. 
\label{Fn-2Fnsimeq}
\end{equation}
The second term vanishes in the limit $i\rightarrow \infty$. 
Substituting (\ref{alpha0=tau-2}) and (\ref{Fn-2Fnsimeq}) into
(\ref{loga=2pifn-2fn-alpha0}), 
\begin{equation}
 \log a =\frac{2\pi \sqrt{5}}{\tau^{2i+1}}
\label{loga2pisqrt5tau2n+1}
\end{equation}
for the orthogonal Fibonacci system $(F_i, F_{i+1})$ with the limit divergence angle (\ref{alpha0=tau-2}). 
This result is generalized to an orthogonal $J$-jugate system $J(F_i, F_{i+1})$ 
with the limit divergence angle $\alpha_0=\tau^{-2}/J$, 
\begin{equation}
 \log a =\frac{2\pi \sqrt{5}}{J\tau^{2i+1}}. 
\label{Jloga2pisqrt5tau2n+1}
\end{equation}
 
%
%


In terms of common logarithms, 
(\ref{loga2pisqrt5tau2n+1}) is expressed as 
\begin{equation}
\log_{10} 
\left(\log_{10} a\right) =
-i {\log_{10} (\tau^{2})}+ 
\log_{10} 
\left(
\frac{2\pi \sqrt{5}}{\tau \log (10)}
\right), 
\end{equation}
which is transformed into 
\begin{equation}
i-1=
\frac{\log_{10} \left(\frac{2\pi \sqrt{5}}{\tau \log (10)}\right)}{
{\log_{10} (\tau^{2})}} -1
- 
\frac{\log_{10} \left(\log_{10} a\right) 
}{\log_{10} (\tau^{2})}. 
\end{equation}
Expressed in numbers, 
\begin{equation}
i-1= 0.37918- 2.39249\log_{10} \left(\log_{10} a\right).  
\end{equation}
The right-hand side is the phyllotaxis index (P.I.) of \cite{richards51}. 
Thus, it is proved that the phyllotaxis index is nothing but 
the index $i$ of the Fibonacci system $(F_i, F_{i+1})$ minus one. 
This is an integer by definition. 
In natural logarithms, the index is given by 
\begin{equation}
{\rm P.I.}  = 
\frac{\log({2\pi \sqrt{5}}/\tau)}{\log \tau^2} 
-\frac{\log (\log a)}{\log \tau^2} -1. 
\label{P.I.log}
\end{equation}
P.I. may take any value if it is regarded as a function of $\log a$. 
In applying the formula (\ref{P.I.log}) to real systems, 
 it should be kept in mind 
that $\log a$ in the logarithm must be a positive number,  namely $a>1$
by assumption. 


The above derivation clearly indicates that
Richards' phyllotaxis index is based on 
the physical assumption (\ref{alpha0=tau-2})
and the mathematical approximation (\ref{Fn-2Fnsimeq}). 
This 
is not obvious from the derivations by \cite{richards51} and \cite{jean83}. 
Interestingly, it is mostly the latter approximation that turns out to be less reliable, i.e., 
the rational number $\frac{F_{i-2}}{F_i}$ for small $i$ cannot be replaced by the irrational number  $\tau^{-2}$. 
The former assumption  (\ref{alpha0=tau-2}), 
or $d\simeq 137.5^\circ$, 
 is empirically supported quite accurately, as shown in the main text. 
In the author's view, 
the regularity expressed by  (\ref{alpha0=tau-2}), or $\theta_n= n d$
with $d=\pm 2\pi/\tau^2$, 
should be called {the fundamental law of phyllotaxis} 
and 
treated particularly as such. 
%
Let us incidentally remark that 
logarithmic spirals of phyllotaxis are far more miraculous than 
similar spirals appearing in certain growing forms  like nautilus shells 
by the very fact that the angle $d$ is not only held constant 
but specifically fixed at $2\pi/\tau^2= 137.5^\circ$. 

For a non-logarithmic spiral pattern, 
parastichy numbers will be shifted in a Fibonacci sequence 
depending on part of the pattern. 
For instance, the square-root spiral by (\ref{rn=Asqrtn}) 
is formally regarded as having the plastochron ratio $a$ depending on the leaf index $n$, namely 
$\log a =  \frac{1}{2(n-1)}\log n$ by $r_n=  a^{n-1}=\sqrt{n}$, 
when a reference scale is set as $r_1=1$. 

%
%
%
%
%
%
%
%



Leaves on a stem 
may be represented in terms of a cylinder coordinate system as 
\begin{eqnarray}
 z_n&=&n h, 
\label{zn=nhthetan=nd}
\\
 \theta_n&=&nd, 
\nonumber
\end{eqnarray}
where $h$ is the length of an internode relative to the girth 
and $d$ is a divergence angle. 
%
Accordingly, the fundamental helix 
is parametrically given by 
\begin{equation}
 z= \frac{b}{2\pi} \theta, 
\label{z=b2pitheta}
\end{equation}
where 
\begin{equation}
 b= \frac{2\pi h}{d}=\frac{h}{\alpha_0}. 
\label{b=2pihd}
\end{equation}
It is easily shown that the angle $\varphi$ that 
the helix (\ref{z=b2pitheta})  makes with the $z$ axis 
is expressed in terms of  $b$ in (\ref{b=2pihd}) 
by the same equation as (\ref{b=cotvarphi}), namely $b=\cot\varphi$, 
Therefore, 
one may follow the same derivation as (\ref{loga=2pisqrt})
to obtain 
\begin{equation}
h = \sqrt{\left(\frac{p}{q}-\alpha_0\right)\left(\alpha_0-\frac{p'}{q'}\right)} 
\label{h=2pisqrt}
\end{equation}
for the orthogonal parastichy system $(q,q')$.  
Note that the plastochron ratio $a$ of a spiral pattern on the apex and 
the internode length $h$ of a cylindrical pattern on the stem 
are formally related by 
\begin{equation}
 h=\frac{1}{2\pi} \log a. 
\end{equation}
Although the cylindrical representation is frequently used in
mathematical studies, 
real systems do not obey (\ref{zn=nhthetan=nd}) 
 because the internode $h$, unlike $d$ and $a$,
 is not constant even approximately. 
(\ref{h=2pisqrt}) should be understood 
as a reference result of an abstract model.

\section{How to number sunflower seeds}
\label{how2number}

To measure divergence angle, 
the sequential order of organs has to be identified according to their age or plastochron. 
Sometimes this may involve laborious tasks, particularly
for a high phyllotaxis pattern like packed seeds on a sunflower head. 
In principle, all seeds reachable from a reference seed along parastichies 
are numbered by successively adding or subtracting parastichy numbers.  
A drawback of this simple method is that a single miscalculation spoils
all the following numbering. 
Therefore, it is practically of much help to have a systematic  device. 
The following method starts from choosing a layer of seeds that lie
near a rim circle, which are regarded as a ``siege'' to start numbering
whole seeds inward.   
In what follows, 
 seeds are counted from the outermost toward the center. 


\begin{table}[t]
\begin{center}
\begin{tabular}{|c|cccccccccc|}
\hline
 $I$  &0 && 1 &&2 &&3 &&4 &\\
\hline
mod($IR,Q$)&0 &$\nearrow$&2  &$\nearrow$&4 &$\searrow$&1&$\nearrow$& 3&$\searrow$ \\ 
\hline
$Q=5$&&2&&2&&1&&2&&1 \\
\hline
\end{tabular}
\caption{
Numbers to make a primordia front circle (see Fig.~\ref{024130}).  
This table is for a parastichy pair of $(Q,Q')=(5,8)$. 
In the middle, mod($IR,Q$) denotes the remainder of $IR$ divided by
 $Q$, 
where $R=2Q-Q'=2$ and an integer $I$ runs from 0 to $Q-1=4$. 
The last row is a 1-2 sequence for $Q=5$, 
which is obtained by positing either 2 or 1 depending on whether the remainder increases or decreases. 
}
\label{table:582}
\end{center}
\end{table}

Let us consider a pattern with opposed parastichies $(Q, Q')$,  where 
integers $Q$ and $Q'$  have no common divisor. 
For definiteness,  $Q$ and $Q'$ are assumed to satisfy 
$1 <Q'/Q <2$, which practically holds true  in most cases. 
(By this assumption, capital letters are used for the parastichy numbers.) 
The bounding integer 2 plays a key role below. 
In addition, an auxiliary integer $R=2Q-Q'$ is introduced. 
For every integer $I=0, 1, 2, \cdots Q-1$, calculate 
the remainder of the division of $IR$ by $Q$, which is denoted as  mod($I R,Q$).  
See Table~\ref{table:582} for $(Q,Q')=(5,8)$ and $R=2Q-Q'=2$. 
In the last line, 
either 1 or 2 is inserted between columns
 depending on whether 
the remainder mod($IR,Q$)  increases or decreases 
as $I$ increases by 1. 
The last 1 at the right bottom is set
 because mod($2 I,5$)$=0$ for $I=0$ on the left end 
is smaller than 3 on the right end for $I=4$. 
There are $R$~ones and $Q-R$~twos, so that the numbers add up to
$R+2(Q-R)=Q'$.  Thus, the 1-2 sequence in  the last line of Table~\ref{table:582}  
represents a partition of $Q' (=8)$ into $Q (=5)$ parts (2+2+1+2+1=8).  

\begin{figure}
\begin{center}
\includegraphics[width=0.7 \textwidth]{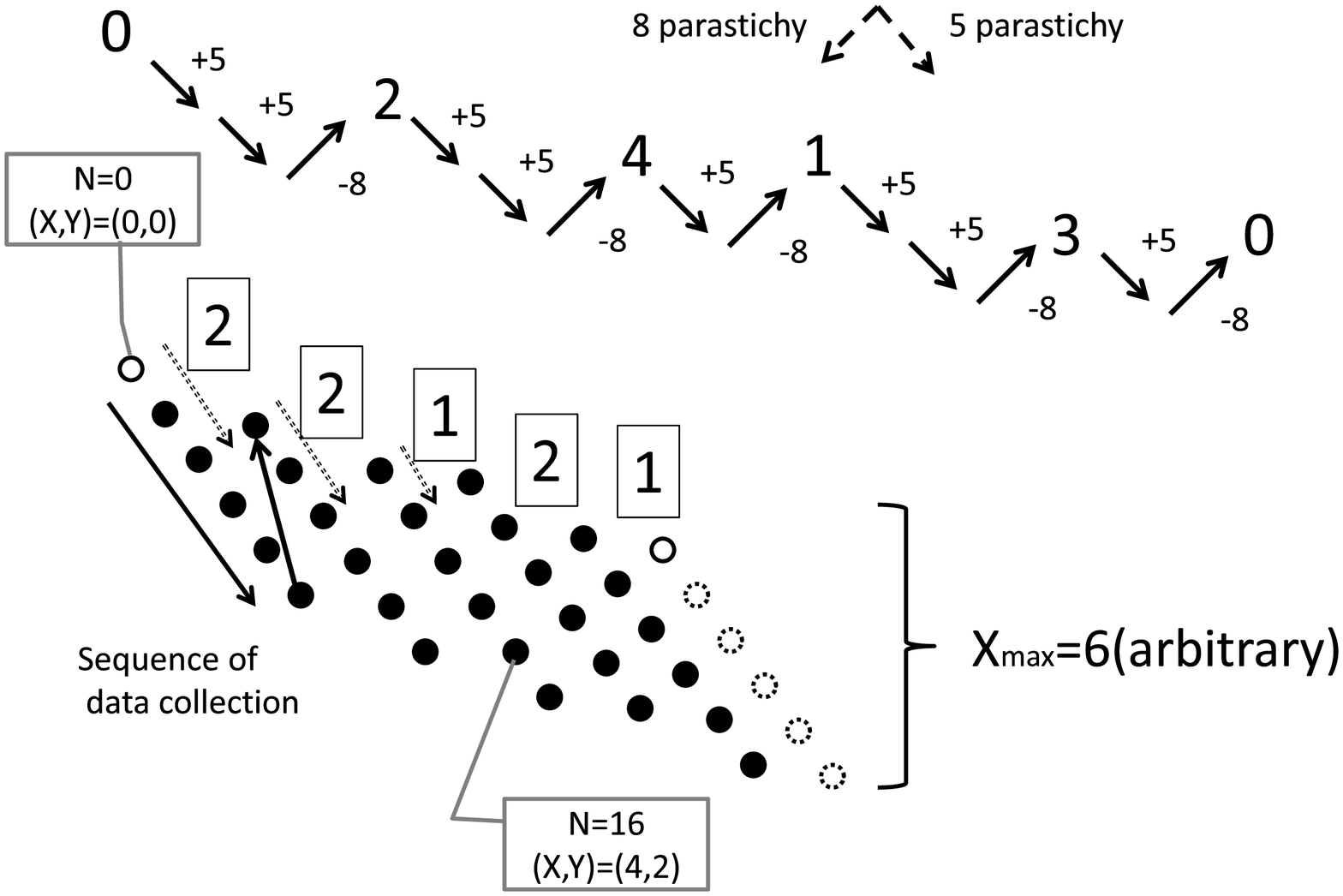}
\includegraphics[width=0.5\textwidth]{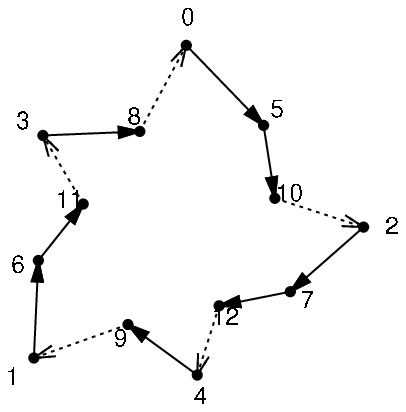}
\caption{How to make an encircling layer of seeds on a parastichy
 lattice with a parastichy pair of $(Q,Q')=(5,8)$. 
See Table~\ref{table:582} and the text.
}
\label{024130}
\end{center}
\end{figure}

With this table at hand, 
Fig.~\ref{024130} explains how to fix 
initial seeds on a front circle 
from which to start numbering the inner seeds. 
The sequence of ones and twos in the last line of Table~\ref{table:582} 
represents the number of steps 
that have to be made in the direction of $Q$ parastichies 
before shifting back one step in the minus direction of $Q'$ parastichies.  
Starting from a reference seed numbered 0, 
the seed 2 is reached
after two steps of $Q$ and minus one step of $Q'$. 
Then it goes to the seed 4 after the same steps, because $2Q-Q'=+2$. 
The next is the seed 1 
by one step along $Q$ and one step back along $Q'$ ($4+Q-Q'=1$). 
In this way, the front circle is closed as 
it returns to the seed 0 after five shifts of $-Q'$.  
The {front circle} thus defined without reference to a nominal center
does not agree with the primordia front of \cite{hjwzagd06}, 
which is defined such that a new primordium added to this front 
lies closer to the center than all primordia belonging to the front.

Cartesian coordinates of seeds may be 
harvested from a digital photo image by
means of a digitizing software (\cite{engauge,digitizer}).  
In so doing, 
useful formulas are derived if one decides to 
pick a constant number of seeds from each of $Q$ parastichies. 
Let this number be denoted as $X_{\rm max}$. 
While collecting 
data in a one-dimensional sequence, 
the starting position in each $Q$ parastichy has to be
successively shifted according to the 1-2 sequence, 
as described in Fig.~\ref{024130}. 
Every seed in the sequential data is assigned 
two-dimensional coordinates $(X,Y)$ set along the $(Q,Q')$ parastichies, 
where $X=0,1,2,\cdots, X_{\rm max}-1$ and $Y=0,1, 2,\cdots, Q-1$. 
In the one-dimensional sequence,  the order of the seed at $(X,Y)$ is given by 
\begin{equation}
 N=X+X_{\rm max} Y, 
\label{N=i+Lj}
\end{equation}
while the plastochron of the seed is 
\begin{equation}
{\rm Plastochron\ Index: }\quad n =
XQ+ {\rm mod}(RY,Q), 
\label{PlastochroneIndex}
\end{equation}
which is not to be confused with the phyllotaxis index in (\ref{P.I.log}). 
(\ref{N=i+Lj}) is inverted to give 
\begin{eqnarray}
 X&=&{\rm mod} (N, X_{\rm max}),
\nonumber\\
Y&=&{\rm int}(N/X_{\rm max}), 
\label{XandY}
\end{eqnarray}
where 
${\rm int}(N/X_{\rm max})= (N-{\rm mod} (N,X_{\rm max}))/X_{\rm max}$ 
means the largest integer that does not exceed $N/X_{\rm max}$. 
By substituting (\ref{XandY}) into (\ref{PlastochroneIndex}), 
the index number $n$ of the $N$-th seed 
is obtained without manual calculation. 
In (\ref{PlastochroneIndex}),  
the initial seed at $(X,Y)=(0,0)$ 
is set to have $n=0$. 
The initial seed is arbitrary but it 
should be chosen properly so as to 
make the front circle as large as possible while keeping the circle within the capitulum.  
Seeds remaining outside the front are indexed with negative numbers. 
Once completed, the index system may be renumbered at one's discretion. 
%

The above method of data collection makes use of the 1-2 sequence. 
If one were interested not in numbering, but only in picking seeds to make a front circle, 
then it is sufficient to know a periodic 1-2 sequence. 
The periodic 1-2 sequence has period $Q$; 
all sequences 22121, 12122, 12212, 21221 and 21212 for $Q=5$ 
are regarded as equivalent. 
A periodic 1-2 sequence may be deduced recursively by simple rules 
to replace 2 with 21 and 1 with 2; 
\begin{eqnarray}
 1&\rightarrow& 2. 
\nonumber\\
 2 &\rightarrow& 21,
\label{rule21}
\end{eqnarray}
For the main sequence of phyllotaxis, starting from 
\begin{equation}
 Q=3:\quad 2\quad 2\quad 1, 
\end{equation}
one gets 
\[
 5:\quad  21\quad 21\quad 2, 
\]
then
\[
 8:\quad  21\ 2\quad 21\ 2\quad 21,  
\]
and so on. 
The rules in (\ref{rule21}) may remind Fibonacci's original problem,  
in which 1 and 2 correspond to new and mature pair of rabbits.  
The 1-2 sequence in the correct order is obtained by 
reversing the sequence before applying (\ref{rule21}). 
It goes as follows: 
\[
 3: 221
\]
\[
122
\]
\[
5: 22121
\]
\[
 12122
\]
\[
8: 22122121
\]
\[
12122122 
\]
\[
13: 2212212122121 
\]
\[
1212212122122 
\]
\[
21: 221221212212212122121 
\]
\[
121221212212212122122 
\]
\[
 34: 2212212122122121221212212212122121 
\] 
The 1-2 sequence depends on $Q'$ too.    
For instance, it is 21211 for $(Q,Q')=(5,7)$, while 22121 for $(Q,Q')=(5,8)$. 
Given a 1-2 sequence for a pair, 
1-2 sequences for other pairs in the same Fibonacci sequence are
routinely derived, according to the rule of reversal and replacement.
From 21211 for $(Q,Q')=(5,7)$,  
the rule gives 2221221 for the next pair $(Q,Q')=(7,12)$ of the $(5,7)$
sequence, 5, 7, 12, 19, $\cdots$. 
The result is confirmed by the method of Table~\ref{table:582}. 






\begin{figure}
  \begin{center}
  \subfigure[]{
\includegraphics[width= .5\textwidth]{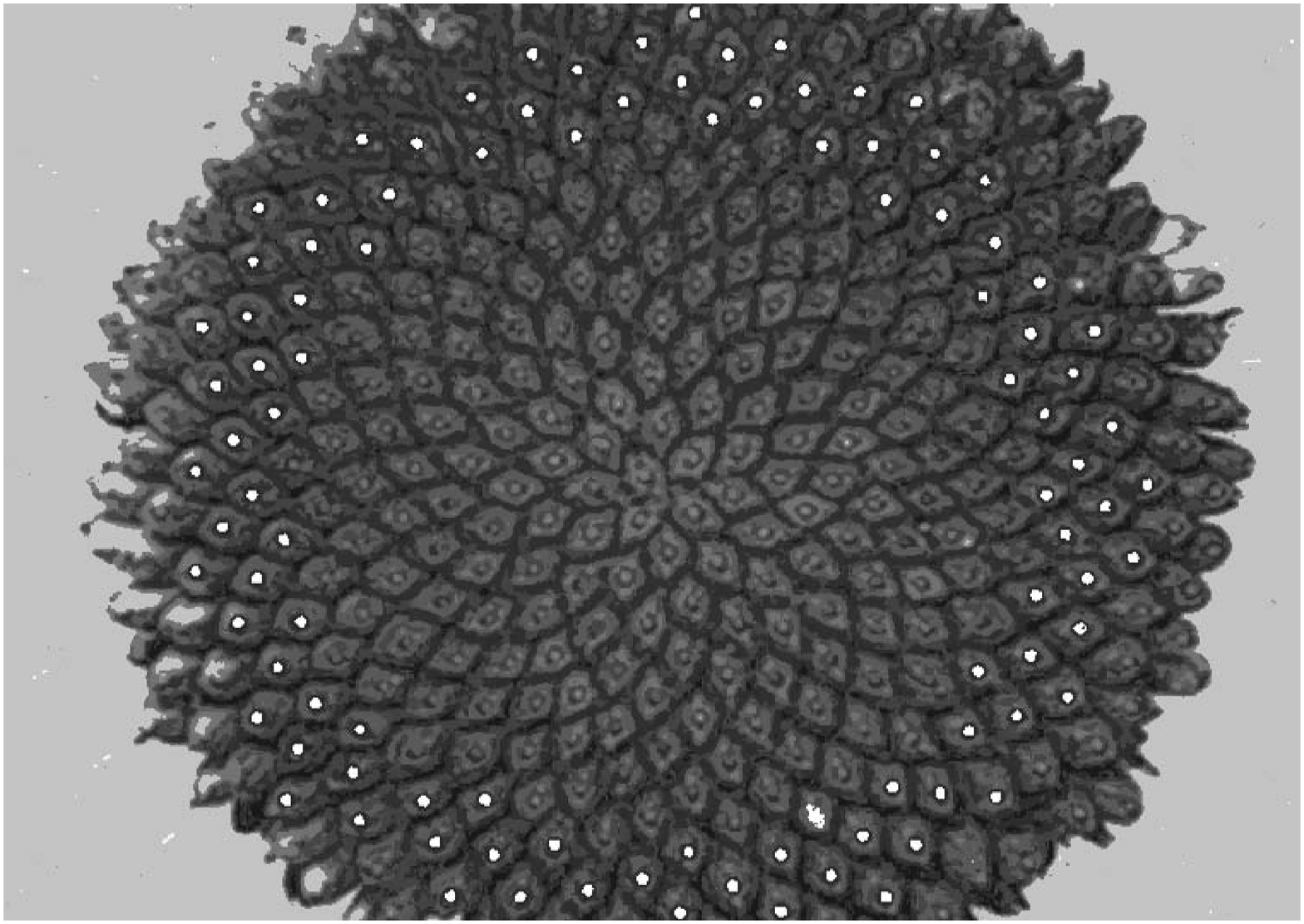}
    \label{jean0}
 }
  \hfill
  \subfigure[]{
\includegraphics[width= .4\textwidth]{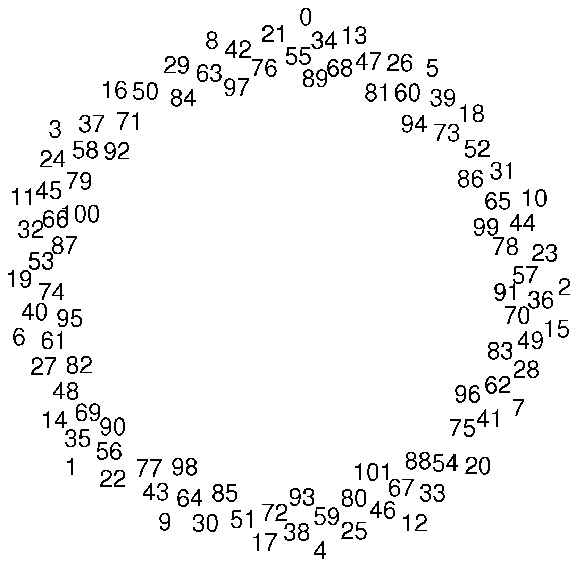}
    \label{jean1}
  }
  \end{center}
  \vspace{-0.5cm}
  \caption{ 
Digitizing an image from \cite{jean94} by way of illustration. 
See the text for a procedure of making the encircling front layer. 
The direction of the fundamental spiral
is opposite to that of Fig.~\ref{realphyllo}. 
}
  \label{jean}
\end{figure}
Returning to the subject, 
let us take the cover image of \cite{jean94} as an example. 
The first thing to do is to detect a shell layer of a
well-defined encircling lattice 
spanned by opposed parastichies. 
In Fig.~\ref{jean0}, there are $Q=34$ main parastichies involuting clockwise
and $Q'=55$ steeper opposite parastichies
($R=13$). 
The ordered 1-2 sequence for $Q=34$ is given just above. 
%
%
For simplicity, let us choose $X_{\rm max}=3$ 
for the thickness of the front layer.  
A procedure is: 
(i) import the image into the digitizer. 
(ii) Set a Cartesian coordinate system with proper scales of $x$ and $y$
axes. 
(iii) Fix
the outermost layer of the primordia front according to the
1-2 sequence. 
Check if the front closes properly (the top and bottom of Fig.~\ref{024130}). 
(iv) Start clicking the seeds in sequence 
(the middle of Fig.~\ref{024130}). 
(v) Index the digitized Cartesian coordinates by  (\ref{PlastochroneIndex}) and (\ref{XandY}). 



The parastichy pair easiest to follow with the eye is to be used as $(Q,Q')$, 
but the choice is not unique. 
The same index system is obtained if a {\it lower} parastichy pair is
used, 
namely $(Q'-Q, Q)$,  $(2Q-Q', Q'-Q)$, etc. 
%
%




Parastichy numbers of a multijugate system 
have a common divisor $J$, 
so that they are expressed as $J(Q,Q')$ 
in terms of 
coprime 
integers $Q$ and $Q'$. 
To close the primordia front of the $J$-jugate system, 
$Q$ steps of a 1-2 sequence 
are repeated $J$ times. 
Meanwhile, the $Y$ coordinate runs through $JQ$ integers 
from 0 to $JQ-1$. 
The plastochron index is given by (\ref{PlastochroneIndex}). 
The jugacy index of 
 the numbering system proposed in Sec.~\ref{sec:whorled} 
shifts orderly along parastichies. 
For the seed at $(X,Y)$, it is given by 
\begin{equation}
{\rm Jugacy\ Index: }\quad 
j={\rm mod}( -PX+P'Y,J) 
\end{equation}
by means of integers $P$ and $P'$ satisfying $PQ'-P'Q=\pm 1$.

%
%
%
%
%
%
%
%
%
%
%



\bibliography{main}

\end{document}